\documentclass[a4,11pt]{article}

\makeatletter
\normalsize
 \newcommand{\bea}{\begin{eqnarray}}
\newcommand{\eea}{\end{eqnarray}}
\newcommand{\be}{\begin{equation}}
\newcommand{\ee}{\end{equation}}
\newcommand{\ba}{\begin{align}}
\newcommand{\ea}{\end{align}}

\newcommand{\Tr}{{\rm {Tr}}}

\newcommand\rref[1]{(\ref{#1})}

\newcommand{\ie}{{\it i.e.}}

\newlength{\slength}
\divide\@tempdima\baselineskip
\@tempcnta=\@tempdima
\skip\@mpfootins = \skip\footins
\renewcommand\section{\@startsection{section}{1}{\z@}%
                                   {-3.5ex \@plus -1.3ex \@minus -.7ex}%
                                   {2.3ex \@plus.4ex \@minus .4ex}%
                                   {\normalfont\large\bfseries}}
\renewcommand\subsection{\@startsection{subsection}{2}{\z@}%
                                   {-2.3ex\@plus -1ex \@minus -.5ex}%
                                   {1.2ex \@plus .3ex \@minus .3ex}%
                                   {\normalfont\normalsize\bfseries}}
\renewcommand\subsubsection{\@startsection{subsubsection}{3}{\z@}%
                                   {-2.3ex\@plus -1ex \@minus -.5ex}%
                                   {1ex \@plus .2ex \@minus .2ex}%
                                   {\normalfont\normalsize\bfseries}}
\renewcommand\paragraph{\@startsection{paragraph}{4}{\z@}%
                                   {1.75ex \@plus1ex \@minus.2ex}%
                                   {-1em}%
                                   {\normalfont\normalsize\bfseries}}
\renewcommand\subparagraph{\@startsection{subparagraph}{5}{\z@}%
                                   {1.75ex \@plus1ex \@minus .2ex}%
                                   {-1em}%
                                   {\normalfont\normalsize\itshape}}

\renewcommand{\@dotsep}{10000}

\def\fnum@figure{\textbf{\figurename\nobreakspace\thefigure}}
\def\fnum@table{\textbf{\tablename\nobreakspace\thetable}}

\long\def\@makecaption#1#2{%
  \vskip\abovecaptionskip
  \sbox\@tempboxa{\small #1. #2}%
  \ifdim \wd\@tempboxa >\hsize
    \small #1. #2\par
  \else
    \global \@minipagefalse
    \hb@xt@\hsize{\hfil\box\@tempboxa\hfil}%
  \fi
  \vskip\belowcaptionskip}

\renewenvironment{thebibliography}[1]{%
\begin{oldthebibliography}{#1}%
\small%
\raggedright%
\setlength{\itemsep}{5pt plus 0.2ex minus 0.05ex}%
}%
{%
\end{oldthebibliography}%
}
\makeatother

\usepackage[colorlinks=true, linkcolor=blue, filecolor=blue, urlcolor=blue, citecolor=blue, anchorcolor=blue, menucolor=blue , linktocpage=true]{hyperref}
\usepackage{scalerel}
\usepackage{graphicx}
\usepackage{amsmath,amssymb}
\usepackage{amsfonts}
\usepackage{physics}
\usepackage{bbold}
\usepackage{mathtools}
\usepackage{verbatim}
\usepackage{cite} 
\usepackage{xcolor}
\usepackage{mdframed}
\usepackage{braket}
\usepackage{enumitem}
\usepackage[normalem]{ulem}

\allowdisplaybreaks
\numberwithin{equation}{section}

\makeatletter
\newcommand{\subalign}[1]{%
  \vcenter{%
    \Let@ \restore@math@cr \default@tag
    \baselineskip\fontdimen10 \scriptfont\tw@
    \advance\baselineskip\fontdimen12 \scriptfont\tw@
    \lineskip\thr@@\fontdimen8 \scriptfont\thr@@
    \lineskiplimit\lineskip
    \ialign{\hfil$\m@th\scriptstyle##$&$\m@th\scriptstyle{}##$\hfil\crcr
      #1\crcr
    }%
  }%
}
\makeatother

\newcommand{\eq}[1]{\begin{equation}#1\end{equation}}

\newcommand{\eqsp}[1]{\begin{equation}\begin{split}#1\end{split}\end{equation}}

\newcommand{\bfO}{{\mathbf{O}}}
\newcommand{\bfOs}{{{\mathbf{O}_{\text{seed}}}}}
\newcommand{\bfOp}{{{\mathbf{O}_{\text{probe}}}}}

\renewcommand{\title}[1]{\vbox{\center\LARGE{#1}}\vspace{5mm}}
\renewcommand{\author}[1]{\vbox{\center#1}\vspace{5mm}}

\newcommand{\email}[1]{\vspace{5mm}\vbox{\center\footnotesize\tt#1}\vspace{5mm}}

\newcommand{\preprint}[1]{\begin{table}[t]  
             \begin{flushright}               
             {#1}                             
             \end{flushright}                 
             \end{table}}

\begin{document}

\preprint{YITP-SB-2025-01}

\thispagestyle{empty}

\phantom{a}
\vskip20mm

\begin{center}

\title{
Symmetric Product Orbifold Universality \\ and the Mirage of an Emergent Spacetime
}

\author{Alexandre Belin$^{\,a,b}$, Suzanne Bintanja$^{\,c,d}$, Alejandra Castro$^{\,e}$, and Waltraut Knop$^{\,f}$
}

\vskip4.1mm

\begin{minipage}[c]{0.89\textwidth}\centering \footnotesize \em {\it $^{a}$}Dipartimento di Fisica, Universit\`a di Milano - Bicocca \\
I-20126 Milano, Italy
\\ \vspace{0.5em}
${}^b$INFN, sezione di Milano-Bicocca, I-20126 Milano, Italy 
\\ \vspace{0.5em} 
{\it $^{c}$Mani L. Bhaumik Institute for Theoretical Physics, Department of Physics and Astronomy,
University of California Los Angeles, Los Angeles, CA 90095, USA}
\\ \vspace{0.5em}
{\it $^{d}$CERN, Theory Division,
Geneva 23, CH-1211, Switzerland}
\\ \vspace{0.5em}
{\it $^{e}$Department of Applied Mathematics and Theoretical Physics, University of Cambridge\\
Cambridge CB3 0WA, United Kingdom}
\\ \vspace{0.5em}
{\it $^{f}$C. N. Yang Institute for Theoretical Physics, Stony Brook University,\\ Stony Brook, NY 11794-3840, U.S.A.}
\\ \vspace{0.5em}
\end{minipage}

\email{alexandre.belin@unimib.it, sbintanja@physics.ucla.edu, ac2553@cam.ac.uk, waltraut.knop@stonybrook.edu}
\end{center}

\vskip6mm

\begin{abstract}

\noindent We study thermal two-point functions and four-point functions involving two heavy twisted operators and two light probes in symmetric product orbifolds. We identify cases where they are universal at large $N$, that is, they are only sensitive to the orbifold structure. Surprisingly, such observables mimic correlators obtained from the BTZ background, even though symmetric product orbifolds are not dual to semi-classical gravity. We discuss the interpretation of these results in light of the criteria for emergence of spacetime via Von Neumann algebras. Our analysis implies that a condition on the infinite $N$ thermal two-point functions cannot be stringent enough to define an emergent spacetime and the concept of a sharp horizon.

\end{abstract}

\eject

\phantom{a}
\vspace{-4em}
{
\tableofcontents
}

\bigskip

\bigskip

\newpage

\section{Introduction}

The holographic nature of quantum gravity implies that the true microscopic degrees of freedom are located on the boundary of spacetime, and the gravitational theory is an emergent description in the bulk of the spacetime. This idea is beautifully realized for quantum gravity with a negative cosmological constant through the AdS/CFT correspondence: theories of quantum gravity in AdS$_{d+1}$ are dual to CFTs in $d$ dimensions, living on the boundary of Anti-de Sitter space. The variables of the bulk theory, as well as its equations of motion, should all be seen as emergent, effective properties that describe some portion of the microscopic dynamics present in the CFT.

For the picture of an emergent spacetime to work accurately, the boundary CFT must be extremely special in the space of all CFTs. Theories of this type are called \textit{holographic} CFTs. While there are many known necessary conditions that CFTs should satisfy in order to be holographic, a set of sufficient conditions is still lacking, i.e., currently there is no smoking gun that can distinguish whether a CFT is holographic or not.

The main strategy that has been pursued so far to remedy our lack of understanding is to identify a set of properties that bulk observables should have, and investigate what they imply on the dual CFT. This has been carried out to great success when the observables are scattering processes on the vacuum AdS background. In the CFT, these map to correlation functions of the CFT in the vacuum, and a program to match bulk and boundary data was initiated in \cite{Heemskerk:2009pn}. The starting point is to consider theories with a large number of degrees of freedom, such that the stress tensor two-point function comes with a large coefficient called $c$, the central charge. From the perspective of the bulk theory, the central charge is related to Newton's constant through
\be
c \sim  \frac{\ell_{\text{AdS}}^{d-1}}{G_N} \,.
\ee
For the bulk theory to be semi-classical, i.e., weakly coupled, Newton's constant must be small in units of the cosmological constant, hence a large central charge is required. The idea of \cite{Heemskerk:2009pn} is to match the couplings of the bulk effective field theory to parameters that appear in the solution to the CFT crossing equation. Ref. \cite{Heemskerk:2009pn} also conjectured that the suppression of higher derivative terms correcting general relativity is tied to the absence of higher spin operators in the dual CFT. At the level of vacuum correlation functions, this has now been firmly established \cite{Afkhami-Jeddi:2016ntf,Meltzer:2017rtf,Belin:2019mnx,Kologlu:2019bco,Caron-Huot:2021enk}.

In this paper, we will be interested in investigating the nature of holographic CFTs, but for thermal observables. These are more interesting because thermal states at sufficiently high temperatures are dual to black holes. Therefore, these thermal observables probe black hole physics. They are, however, also more complicated. The bootstrap for thermal correlators and its holographic consequences is still being developed, see for example \cite{El-Showk:2011yvt,Iliesiu:2018fao,Fitzpatrick:2019zqz,Fitzpatrick:2020yjb,Alday:2020eua,Parnachev:2020fna,Karlsson:2022osn,Marchetto:2023xap}. 

In two dimensions the situation is improved due to modular invariance, and powerful results have been obtained. Hartman, Keller, and Stoica (HKS) derived the condition that two-dimensional CFTs should satisfy to match the thermal phase structure of AdS$_3$ gravity \cite{Hartman:2014oaa}. The result is a condition on the light spectrum known as a sparseness condition: in the limit $c\to \infty$, 
\be
 \rho(\Delta) \lesssim e^{2 \pi \Delta } \,, \quad  \Delta < \frac{c}{12}  \qquad \Longleftrightarrow \qquad Z_{\text{CFT}}(\beta) =  Z_{\text{grav}}(\beta)=\begin{cases}e^{\frac{c}{12}\beta} \quad &\beta>2\pi \\
e^{\frac{c}{12}\frac{4\pi^2}{\beta} } \quad &\beta<2\pi  \end{cases} \,.
\ee
For a CFT$_2$ that satisfies this sparseness condition, the entropy of the CFT$_2$ matches the Bekenstein-Hawking formula when the BTZ black hole dominates the canonical ensemble, and just like AdS$_3$ gravity, exhibits a Hawking-Page phase transition at the self-dual temperature.\footnote{These results have now been extended in full generality, also with angular potential in \cite{Dey:2024nje}.} Under certain growth conditions of the OPE coefficients, these results were also extended to correlation functions in \cite{Kraus:2017kyl}. Still, these results are far from a sufficient condition for holographic CFTs. To see this, note that there are known CFTs that satisfy (and in fact saturate) the sparseness bound, yet are known not to be holographic. A prime example, which forms the main protagonist of this paper, is the class of CFTs known as symmetric product orbifolds \cite{Keller:2011xi}. This demonstrates that the constraints coming from thermal physics must somehow supplement and not replace those already known from vacuum correlators. 

\subsection*{Operator algebras as a smoking gun?}

A new perspective on the emergence of spacetime was put forward in \cite{Leutheusser:2021qhd,Leutheusser:2021frk}. This proposal comes from a different set of ideas, tied to algebraic quantum field theory and operator algebras, rather than the conformal bootstrap. The main concept is that the operator algebras of holographic CFTs in thermal states above the Hawking-Page phase transition have a distinguished feature: they go from a type I von Neumann algebra at finite $N$ to a type III$_1$ von Neumann algebra at infinite $N$. Type I von Neumann algebras are the standard encountered in quantum mechanics when the spectrum is discrete. CFTs on compact manifolds have a discrete spectrum so they are clearly of type I. Type III$_1$ von Neumann algebras, on the other hand, arise in quantum field theory, where the spectrum of the modular Hamiltonian of a subregion is always continuous, and traces like the one involved in entanglement entropy are not well defined. In the holographic context, this continuous spectrum is associated with the black hole horizon. Any probe field propagating on a black hole background has a continuous spectrum. This jump in the Von Neumann algebra was proposed by \cite{Leutheusser:2021qhd,Leutheusser:2021frk} as a smoking gun for the appearance of a horizon, and thus of an emergent spacetime description. Many developments have followed, going from $1/N$ corrections and generalized entropies \cite{Witten:2021unn,Kudler-Flam:2023qfl} to local operator algebras in quantum gravity \cite{Bahiru:2022oas,Bahiru:2023zlc,Jensen:2023yxy} and a connection with information loss \cite{Furuya:2023fei}, and to de Sitter space \cite{Chandrasekaran:2022cip,Witten:2023qsv,Chen:2024rpx,Kudler-Flam:2024psh}, among others.

Von Neumann algebras in the context of holography are best understood for theories dual to Einstein gravity. What happens at finite string length is more mysterious. For example, what type of operator algebra appears in $\mathcal{N}=4$ at infinite $N$ but at finite $\lambda$? One thing is known for sure: in the free theory, i.e. at $\lambda=0$, the algebra is of type I. This follows from the fact that the spectrum of the free theory is half-integer spaced, so even above the deconfinement phase transition, the two-point function is a periodic function with period $4\pi$. The structure of the operator algebra and the meaning of the horizon at finite $\lambda$ was discussed in \cite{Gesteau:2024rpt}. There, from algebraic considerations at infinite $N$, the authors define the so-called causal depth parameter; a quantity that measures the depth of the emergent radial direction in the bulk as seen by a probe field. An infinite causal depth parameter for all light probe operators is proposed as a sign for an emergent description in terms of a local spacetime with a bifurcate horizon, whereas having differing values for the causal depth parameter points to a stringy geometry. In this case, the equivalence principle is broken, and different probe fields experience different geometries.

In this paper, we will test some of these ideas in the context of AdS$_3$/CFT$_2$. In particular, we will focus on symmetric product orbifolds.  This is a class of two-dimensional CFTs with a well-defined large-$N$ limit. They are constructed by considering a seed CFT $\mathcal{C}$, taking the $N$-fold tensor product of $\mathcal{C}$, and gauging the global $S_N$ symmetry that acts by permuting the different tensor factors
\eq{
\text{Sym}^N(\mathcal{C})=\frac{\mathcal{C}^{\otimes N}}{S_N}\,.
}
The central charge of the symmetric product orbifold is equal to $c=c_0N$, with $c_0$ the central charge of the seed theory. Symmetric product orbifolds are known to obey many properties of holographic CFTs: they have a sparse spectrum, and thus a universal thermal partition function. Moreover, they satisfy large-$N$ factorization. On the other hand, they always have higher spin currents, have a Hagedorn growth of light states, and are not chaotic. These properties are a consequence of the fact that symmetric product orbifolds are free discrete gauge theories, and hence are not strongly coupled, which is expected for a theory dual to weakly coupled Einstein gravity. One should view them as the AdS$_3$/CFT$_2$ version of free $\mathcal{N}=4$ SYM. Indeed, if the seed theory is chosen to be the non-linear sigma model on $T^4$ or $K3$, the bulk dual is known to be a tensionless string theory in AdS$_3$ \cite{Eberhardt:2018ouy,Eberhardt:2019ywk}. 

Concretely, we will study thermal correlation functions of symmetric product orbifolds in the large-$N$ limit. Our main result is that thermal correlators of symmetric product orbifolds, above the Hawking-Page transition, take the exact same form at large $N$ as bulk correlators in the BTZ geometry. For example, for the Euclidean two-point function at finite temperature, we find the following universal result for untwisted sector operators
\eqsp{
\left\langle\bfO(t_\text{E},\phi)\bfO(0)\right\rangle_\beta \underset{N\to\infty}{\to}
\sum_{j\in\mathbb{Z}}\frac{1}{\left(\frac{\beta^2}{4\pi^2}\left(\cosh\left(\frac{2\pi(\phi+2\pi j)}{\beta}\right)\right)-\cos\left(\frac{2\pi t_\text{E}}{\beta}\right)\right)^{2h_\bfO}} \,, \quad \text{for}\quad \beta < 2 \pi \,.
}

This result has multiple consequences. First, it means that in top-down examples such as the D1-D5 CFT where the symmetric product orbifold is a point on moduli space,  protected operators have the same thermal two-point function at any value of the coupling to leading order in the large-$N$ limit. It is important to emphasize that thermal correlators are not protected, and it is remarkable that the thermal correlators of these operators do not change as a function of the coupling. Second, turning to the operator algebra perspective, it means that the operator algebra associated to any single-trace operator of the CFT is the \textit{exact same} operator algebra as that of the strongly coupled case. There, the bulk geometry takes the form of the BTZ black hole, and the horizon makes the operator algebra type III$_1$. This should be sharply contrasted with the case of $\mathcal{N}=4$ SYM at zero coupling, where the spectrum is half-integer spaced and the algebra remains type I even above the deconfinement transition.

Our results challenge the idea that the type of Von Neumann algebra that appears at large $N$ in thermal states is a sufficient condition for the emergence of a geometric spacetime. As we already mentioned, symmetric product orbifolds capture the features of a tensionless string theory in AdS$_3$. These bulk theories are in some sense as different as possible from Einstein gravity: they contain infinite towers of massless higher spin operators and do not display Lyapunov growth in out-of-time-ordered correlators. One possibility, which we expand on in the discussion, is that the operator algebra of symmetric product orbifolds is actually not type III$_1$ at large $N$, even though the subalgebra generated by each single-trace operator itself is type III$_1$. This breakdown would come from resuming the exponentially growing number of single-trace operators due to stringy degrees of freedom. This idea is compatible with recent calculations of entanglement entropy in string theory where a finite answer is obtained, indicating that the operator algebra of free string theory is of type I \cite{Dabholkar:2023ows,Dabholkar:2023tzd}.

We also compute two-point functions of light operators in other ensembles of states, like ground states of twisted sectors; in the supersymmetric top-down examples, the ensemble of twisted ground states are half BPS Ramond-Ramond ground states and are dual to microstates of a black hole with a classically vanishing area (see e.g. \cite{Balasubramanian:2005qu}). We find a large amount of universality in these correlators as well: for all the probes we consider, the two-point functions are universal. This does not mean that the two-point functions are universal in all microstates (which would be essentially equivalent to the statement that symmetric product orbifolds satisfy the Eigenstate Thermalization Hypothesis): there exist particular microstates where universality breaks down. But these microstates are highly atypical, and the vast majority of microstates have in fact universal correlators. The key ingredient needed is the existence of cycles that scale with $N$ in the twisted sectors under consideration, which the majority of sectors have.\footnote{Along the way, we fix a statement presented in \cite{Balasubramanian:2005qu} concerning typical twisted sectors; see Sec.\,\ref{sec:twisteduniv} for further details.}

This paper is organized as follows. In Sec.\,\ref{sec:rev}, we review the construction of symmetric product orbifolds. We discuss properties that these theories share with holographic CFTs, and also ones that tell them apart. In Sec.\,\ref{sec:thermaluniv}, we study thermal correlators in symmetric product orbifolds. We write down the thermal correlators in closed form, and discuss the properties of such correlators at large $N$. In Sec.\,\ref{sec:twisteduniv}, we study correlation functions in twisted ground states of the symmetric product orbifold and find again universality in the large-$N$ limit. We conclude with a discussion on the gravitational interpretation of our results and future directions in Sec.\,\ref{sec:grav}. We include a review of the dominant contribution to the thermal partition function of symmetric product orbifolds in App.\,\ref{app:longcycles}, and we review the complementary approach to thermal correlation functions based on the sparseness conditions of \cite{Kraus:2017kyl} in App.\,\ref{app:hks2pt}.

\section{A brief review of symmetric product orbifolds} \label{sec:rev}

In this section, we review some salient features of symmetric product orbifolds. We first discuss their defining properties and then review which aspects of them are known to be universal in the large-$N$ limit. We also discuss their implications for a holographic interpretation in terms of an emergent spacetime. 

\subsection{General properties} \label{sec:symN}

The construction of symmetric product orbifolds consists of two steps. First, we start with a seed CFT $\mathcal{C}$, which we take to be compact. Often, the seed CFT is taken to be a theory that is under good mathematical control.\footnote{Notable examples that have appeared in the context of holography are nonlinear sigma models or minimal models.} The reason for this is that the observables of the orbifold theory are fully determined in terms of those of the seed theory, and if the seed theory is solvable, then so is the orbifold. Nevertheless, for the purpose of this paper, the seed can be any compact CFT. The structure we present below is universally valid and does not depend on the choice of seed. The idea is to first take a tensor product of $N$ copies of the seed $\mathcal{C}$: 
\be
\mathcal{C} \to \mathcal{C}^{\otimes N} \,.
\ee 
This introduces a global $S_N$ symmetry that acts by permuting the different copies in the tensor product.  

The second step is then to gauge the global $S_N$ symmetry.\footnote{One could also choose to not gauge by the entire global symmetry, but by some subgroup $G_N\subseteq S_N$. The resulting theories are called permutation orbifolds and have been studied in e.g. \cite{Klemm:1990df,Haehl:2014yla,Belin:2014fna,Belin:2015hwa,Belin:2017jli,Keller:2017rtk}.} This has two effects: first, it removes all the operators that are not invariant under $S_N$. This is a projection to the singlet sector. But the resulting theory is not modular invariant, and hence the second effect is to add other operators: the twisted sectors. An operator in the twisted sector is characterized by a nontrivial boundary condition labeled by a group element $g\in S_N$. More concretely, an operator $\bfO$ in the twist $g$ sector satisfies 
\eq{
\bfO^{(i)}\left(e^{2\pi i}z\right)=\bfO^{(g(i))}(z)\,,
}
where the superscript denotes the copy on which $\bfO$ is inserted. We will denote such twist-$g$ operators by $\bfO_{[g]}$.
The resulting Hilbert space thus consists of symmetrized seed states that form the untwisted sector and symmetrized twisted states, which can be labeled by the conjugacy classes of $S_N$. The resulting theory is called the symmetric product orbifold of $\mathcal{C}$, and is denoted by
\eq{
\text{Sym}^N(\mathcal{C})\coloneqq \frac{\mathcal{C}^{\otimes N}}{S_N}\,.
}
The central charge of the symmetric product orbifold theory is enhanced by a factor $N$ compared to the seed central charge $c_0$
\eq{
c=Nc_0\,.
}

As mentioned above, a major upside of considering symmetric product orbifolds is that any observable in the orbifold theory can be computed in terms of seed data (typically products of observables of the seed theory). Therefore, this class of theories gives rise to tractable families of large-$N$ CFTs that are under good mathematical control. 

\subsubsection{Correlation functions}\label{sec:RevCorr}

Let us briefly discuss the computation of correlation functions in symmetric product orbifolds, since the main goal of this paper is to understand universality in correlation functions. In this section we discuss the methods we use to compute such correlation functions. Explicit examples will be considered in the following sections. Untwisted sector operators are directly inherited from the seed CFT, hence correlation functions on the plane of (normalized) untwisted operators can be found by computing seed correlators and adding the appropriate combinatorial factors. Including twisted sector insertions in correlation functions is more intricate, since the presence of a twisted sector operator modifies the periodicity of operators around its insertion. We follow the method of \cite{Lunin:2000yv,Lunin:2001pw} to compute correlation functions that involve twisted-sector operators.

Recall that twisted operators are labeled by a conjugacy class of $S_N$. Concretely, that means that the twisted sectors are characterized by the cycle structure of the permutations that represent the conjugacy class, which are in one-to-one correspondence with partitions of $N$. For simplicity let us restrict to twisted operators that come from a conjugacy class consisting of a single cycle of length $w$, which we will denote by $\sigma_{w}$. Generalizations can be obtained by taking coincidence limits of these single-cycle insertions. Say we want to compute the following $n$-point correlation function
\eq{
\langle \sigma_{w_1}(z_1)\cdots\sigma_{w_n}(z_n)\rangle_X\,,
}
on some manifold $X$. The idea is then to decompose the twisted insertions as a sum over the different choices of representatives of the conjugacy classes, and compute the sum of the correlators of these non-gauge invariant operators. Terms in the sum look like
\eq{\label{eq:twistgcor}
\langle \sigma_{g_1}(z_1)\cdots\sigma_{g_n}(z_n)\rangle_X\,,
}
where $g_i\in S_N$ is given by a single cycle of length $w_i$. Many of the terms in the sum vanish, since the permutations involved have to multiply to the identity to give a nonzero result
\eq{\label{eq:id}
g_1\cdots g_n=\text{id}\,.
}
The physical intuition is that by starting on say the first copy and cycling around all twist insertions one has to return to the first copy again to have a well-defined monodromy.

The trick to compute correlators of the form \eqref{eq:twistgcor} is to move to a cover space $\Sigma$ where the fundamental fields of the seed theory are single-valued, and the twisting gets undone. The key to understanding these cover spaces is the so-called cover map $\Gamma: \Sigma\rightarrow X$. The cover map has the following behavior around the twisted insertions
\eq{\label{eq:covermap}
\Gamma(t)\sim z_i+b_{t_i}^\Gamma (t-t_i)^{w_i}+\cdots\,.
}
Here $b_i^\Gamma$ are constants, and the $t_i$ are the points on $\Sigma$ over which $\Gamma$ is branched with order $w_i$. This behavior ensures that going $w_i$ times around $t_i$  corresponds to one full rotation around its pre-image $z_i$, the correct periodicity for a twist-$w_i$ operator.
Hence, on the cover surface \eqref{eq:twistgcor} reduces to a correlation function of (single-valued) seed insertions. The correlator \eqref{eq:twistgcor} is then given by summing over all possible cover surfaces, that is, all surfaces that allow for a cover map $\Gamma$ satisfying \eqref{eq:covermap}. 

Note that the cover surface generically is not a connected surface. For example, if there are copies of the symmetric product orbifold not involved in the $g_i$, the cover map restricted to those copies can be taken as the identity map. Connected components of the cover space correspond to the pre-image sets of $n_c$ copies for which the twist-insertions involving those copies form a transitive subgroup of $S_{n_c}$.

\subsubsection{The partition function}\label{sec:z}

A striking example of the simplicity with which orbifold quantities can be computed in terms of seed data is the partition function. Let $Z_\mathcal{C}$ be the partition function of the seed
\eq{
Z_\mathcal{C}(\tau,\bar\tau)\coloneqq\Tr\left[q^{L_0-c/24}\bar{q}^{\bar{L}_0-c/24}\right]=\sum_{n,\bar{n}} d(n,\bar{n})q^n\bar{q}^{\bar{n}}\,,\qquad\text{where}\qquad q\coloneqq e^{2\pi i \tau}\,.
}
In symmetric product orbifolds, it is often convenient to work in the grand canonical ensemble, where we introduce a chemical potential conjugate to the central charge
\be
p= e^{2 \pi i \rho  } \,.
\ee
The grand canonical partition function is thus the generating function for the orbifold partition functions. It can be written in extremely compact form \cite{Dijkgraaf:1996xw,Bantay:2000eq}:
\eq{\label{eq:ZHecke}
\mathcal{Z}(\tau,\bar{\tau},\rho)\coloneqq \sum_{N=0}^\infty p^N Z_{S_N}(\tau,\bar{\tau})=\text{exp}\left(\sum_{L=0}^\infty \frac{p^L}{L}T_{L} Z_\mathcal{C}(\tau,\bar{\tau})\right)\,,
}
where $T_L$ is the $L$-th Hecke operator which acts like
\eq{
T_L f(\tau,\bar{\tau})=\sum_{\substack{a,d=1\\ ad=L}}^L\sum_{b=0}^{d-1} f\left(\frac{a\tau+b}{d},\frac{a\bar{\tau}+b}{d}\right)\,.
}
Furthermore, \cite{Dijkgraaf:1996xw} showed that we can write the grand canonical partition function in terms of the seed density of states as
\eq{\label{eq:dmvv}
\mathcal{Z}(\tau,\bar{\tau},\rho)=\prod_{m>0}\prod_{\substack{n,\bar{n}\\n-\bar{n}=0\text{\ mod\ }m}}\left(1-p^mq^{n/m}\bar{q}^{\bar{n}/m}\right)^{-d(n,\bar{n})}\,,
}
which is known as the DMVV formula (sometimes also referred to as the product formula). The partition function of the $N$-th symmetric product orbifold can thus be found by expanding out the products in the DMVV formula and extracting the coefficient multiplying $p^N$.

At a given value of $N$, there exists another formula for the partition function of the symmetric product orbifold, which is more group-theoretic based. It is known as Bantay's formula \cite{Bantay:1997ek,Bantay:1999us}:
\be
Z_{S_N}(\tau,\bar{\tau})=\frac{1}{\abs{S_N}}\sum_{\substack{g,h\in S_N\\gh=hg}}\prod_{\xi\in O(g,h)}Z_\mathcal{C}(\tau_\xi,\bar{\tau}_\xi)\,.
\ee
The advantage of Bantay's formula is that being based on group theory, it applies in a straightforward manner to any permutation orbifold, as only the orbits and subgroup structure will change between different permutation groups. Let us discuss in more detail the different elements that appear in Bantay's formula and generalize to any permutation group $G_N\subseteq S_N$.

In order to understand the contributions from twisted sectors to the torus partition function, we have to consider allowed cover surfaces of the torus. In the case of a permutation orbifold by permutation group $G_N$, these are characterized by homomorphisms of the fundamental group of the torus, $\mathbb{Z}\oplus\mathbb{Z}$, into $G_N$. This boils down to a pair of commuting elements in the permutation group that we will denote by $g,h\in G_N$ (where as mentioned $gh=hg$). Concretely, $g$ and $h$ encode the nontrivial homology around the two cycles of the torus. In other words, they denote how the different copies of the seed are permuted once you go around the different cycles of the torus. Generically the resulting cover surface has different connected components, each of which is again a torus due to the Riemann-Hurwitz formula. The connected components are in one-to-one correspondence with the orbits of the abelian group generated by $g$ and $h$ acting naturally on the set $\{1,\dots,N\}$. We denote the set of orbits by $O(g,h)$. An orbit $\xi\in O(g,h)$ can then be characterized by three integers
\begin{enumerate}
    \item $\lambda_{\xi}$: the length of $g$ orbits inside $\xi$,
    \item $\mu_\xi$: the number of $g$ orbits in $\xi$,
    \item $\kappa_\xi$: the smallest nonnegative integer such that $g^{\kappa_\xi}=h^{\mu_\xi}$ on all points in $\xi$.
\end{enumerate}
The modular parameter of the torus associated to $\xi$ is given by
\eq{\label{eq:tauxi}
\tau_\xi=\frac{\mu_\xi\tau+\kappa_\xi}{\lambda_\xi}\,.
}
The torus partition function of a permutation orbifold is then equal to the sum of the seed partition function on the tori making up all allowed coverings of the torus under consideration, normalized appropriately, i.e.,
\eq{\label{eq:bantay}
Z_{G_N}(\tau)=\frac{1}{\abs{G_N}}\sum_{gh=hg}\prod_{\xi\in O(g,h)}Z_\mathcal{C}(\tau_\xi)\,.
}
Here the sum takes into account all allowed cover surfaces, while the factors appearing in the product account for the seed partition function on the connected components of the corresponding cover surface. 

The terms in the sum with $g=(1)$ give rise to the untwisted sector partition function, as shown in \cite{Belin:2014fna}. We can thus keep the following intuition in mind. Nontrivial choices for $g$ correspond to picking out different twisted sectors, while the sum over $h$ takes care of the symmetrization over the permutation group. For the case of the symmetric group, i.e. $G_N=S_N$, \eqref{eq:bantay} reproduces the relevant term in the sum over Hecke operators in the generating function for symmetric product orbifolds \eqref{eq:ZHecke}, see \cite{Bantay:2000eq}.

\subsection[Universality at large \texorpdfstring{$N$}{N}]{Universality at large $\boldsymbol{N}$} \label{sec:symuniversality}

We are interested in understanding the properties of symmetric product orbifolds in the large-$N$ limit, and to what extent universality arises. As we will see, there are properties of symmetric product orbifold theories that arise in the $N\rightarrow\infty$ limit that are completely universal, i.e., they are true irrespective of the choice of seed theory. Various universal quantities and properties have been discovered in the past, and in this section we provide an overview of these properties. The rest of the paper will be devoted to extending the list of universal properties.

We will organize our review according to whether the universal properties support a standard emergent spacetime (i.e., Einstein's gravity weakly coupled to local matter degrees of freedom) or not. The take-home message of this section is that even though symmetric product orbifolds have certain advantageous universal properties in light of a holographic interpretation, there are also universal facts that spoil a potential duality with weakly coupled Einstein gravity, as we will see momentarily.

\subsubsection{Similarities with low-energy gravity}

We start by reviewing universal properties of symmetric product orbifolds that support a gravitational interpretation in terms of general relativity weakly coupled to a local QFT. In particular, we will describe the universal phase structure of the thermal partition function and its associated density of states, as well as large-$N$ factorization.

\paragraph{Phase structure and entropy} An important requirement for a holographic interpretation of any CFT is that it should reproduce the thermodynamics of black holes. This means that it should display the same phase structure as gravity, and the CFT density of states, in the regime where black holes dominate, should agree with the Bekenstein-Hawking formula. In the context of two-dimensional CFTs this requirement is particularly constrained due to modular invariance. The phase structure of gravity in AdS$_3$ is as follows: at low temperatures, thermal AdS is the dominant geometry while at high temperatures it is the BTZ black hole. They exchange dominance at the temperature $\beta=2\pi$ which is self-dual under the S-transformation of the modular group. The phase structure is thus
\be \label{phase3dgrav}
Z_{\text{grav}}(\beta) = \begin{cases} e^{\frac{c}{12} \beta }  \qquad & \beta> 2 \pi \\
e^{\frac{c}{12} \frac{4\pi^2}{\beta } } \qquad & \beta< 2 \pi
\end{cases} \,.
\ee
At the self-dual temperature, the entropy jumps and this describes a first-order phase transition: the Hawking-Page phase transition. In the high-temperature phase, the density of states is given by the Bekenstein-Hawking formula of the BTZ black hole, which in CFT language gives
\be
\rho(E) = e^{2\pi \sqrt{\frac{c}{3} E} } \,,
\ee
which is valid as soon as the BTZ black hole dominates the thermal ensemble, namely
\be
 E \geq \frac{c}{12} \,.
\ee
Note that this density of states is nothing more than the Cardy formula, but it is \textit{not} the usual Cardy regime which would require $c$ fixed as $E\to\infty$. Here, we have an extended regime of validity of the Cardy formula and we can trust it in the regime
\be
E \sim c \,, \qquad c \to \infty \,.
\ee

Let us now return to symmetric product orbifolds. What is the phase diagram of symmetric product orbifolds at large $N$? It turns out to be exactly the same as gravity, for any choice of seed CFT! We thus have
\be
Z_{S_N}(\beta) = \begin{cases} e^{\frac{c}{12} \beta }  \qquad & \beta> 2 \pi \\
e^{\frac{c}{12} \frac{4\pi^2}{\beta } } \qquad & \beta< 2 \pi
\end{cases} \,,
\ee
for any choice of $\mathcal{C}$.

This result has been derived in two ways. First, it was derived by direct inspection of the product formula \rref{eq:dmvv} in the large-$N$ limit \cite{Keller:2011xi}. We will review some aspects of this derivation in App.\,\ref{app:longcycles}. There is also a second way to derive this result. Independently of symmetric product orbifolds, one can ask the question: since modular invariance relates the high and low spectrum, under which conditions on the light spectrum does one obtain the phase structure \rref{phase3dgrav} along with the associated extended regime of validity of the Cardy formula. This question was studied by Hartman, Keller, and Stoica (HKS) who showed that if the spectrum satisfies the following sparseness condition 
\eq{\label{eq:HKS}
\rho(\Delta)\lesssim e^{2\pi\Delta}\,,  \qquad \text{for}\qquad \Delta < \frac{c}{12} \,,
}
where $\Delta$ denotes the conformal dimension of operators in the spectrum, the theory exhibits a phase transition at $\beta=2\pi$ \cite{Hartman:2014oaa}, and the phase structure is the same as that of 3D gravity. Symmetric product orbifolds turn out to obey the HKS bound \rref{eq:HKS} for any choice of seed theory \cite{Keller:2011xi}.\footnote{In fact, they saturate it, which turns out not to be a desirable effect for holographic purposes as we will discuss below.}

The upshot is that symmetric product orbifolds have a thermal partition function that is indistinguishable from that of 3D gravity in the large-$N$ limit, which makes them very interesting for holographic purposes.\footnote{It is worth mentioning that the $1/N$ corrections to the partition function will not agree with those expected of a gravitational theory.}

\paragraph{Large-$\boldsymbol{N}$ factorization.}

There is another universal property of symmetric product orbifolds in the large-$N$ limit that impacts correlation functions. It goes under the name of large-$N$ factorization. In large-$N$ gauge theories, light operators, i.e., operators whose scaling dimensions are finite in the $N\rightarrow\infty$ limit, can be organized according to their number of traces. The $N$-scaling of their correlation functions depends critically on whether the operators have a single trace or more than one trace.

In the case of symmetric product orbifolds the number of traces corresponds to the number of independent sums necessary to symmetrize the seed operator insertions on the different copies. A single-trace operator will thus be either an untwisted sector operator where only a single copy is excited, or a twisted sector with a conjugacy class given by a single cycle. All other light operators, namely twisted-sector operators with multiple cycles, untwisted insertions on multiple copies, or a combination of cycles and untwisted operator insertions all correspond to multi-trace operators.

This terminology in terms of traces is really only useful if the single-trace operators ($\bfO^{\text{st}}_i$) behave as generalized free fields, i.e., the connected contribution to correlation functions of single-trace operators is suppressed in powers of $1/N$
\eq{\label{eq:factorization}
\left\langle\bfO^{\text{st}}_1\cdots \bfO^{\text{st}}_n\right\rangle_{\text{connected}}\sim N^{-(n-2)/2}\,.
}
Here operators are normalized such that the two-point function is $\langle\bfO_i(0)\bfO_j(1)\rangle=\delta_{ij}$.
It follows that in the large-$N$ limit, all the non-vanishing contributions to correlation functions of light operators can be found via Wick contractions. For single-trace operators, these are always disconnected contributions. Correlations of multi-trace operators on the other hand can have a connected piece at order $N^0$.

Large-$N$ factorization thus provides an organizational principle for correlation functions of light operators. On the gravity side, large-$N$ factorization implies that the gravitational description has a weak coupling limit \cite{El-Showk:2011yvt}. Since $N$ is related to Newton's constant $N\sim\frac{1}{G_N}$, large-$N$ factorization implies that all bulk higher-point interactions are suppressed by powers of $G_N$ in the semiclassical limit. 

Symmetric product orbifolds satisfy large-$N$ factorization due to the inherited combinatorics \cite{Pakman:2009zz}. They thus display the same large $N$ structure as gauge theories and have an associated genus expansion. In the case of large-$N$ adjoint gauge theories, this follows from properties of double-line Feynman diagrams. Here, it follows from combinatorics and properties of cover maps. Large-$N$ factorization also applies to a large class of permutation orbifolds \cite{Belin:2015hwa,Belin:2017nze}.

\subsubsection{Differences from low-energy gravity}

We have just seen that symmetric product orbifolds have multiple appealing universal properties in light of a holographic interpretation. However, they are never dual to standard low-energy bulk duals, namely Einstein gravity with local matter fields. This is because they have other universal properties, which are in tension with semi-classical aspects of general relativity coupled to matter. Here, we list some of the properties that give rise to the tension.

\paragraph{Higher spin currents.}

Due to the structure of symmetric product orbifolds in terms of a $N$-fold tensor power gauged by the symmetric group $S_N$, the orbifold theory has a greatly enhanced chiral algebra. This can be seen from the fact that any permutation invariant combination of the currents of the seed theory forms a current in the orbifold theory, usually of higher spin. The simplest higher spin current that is a consequence of the orbifold structure can be constructed using the stress tensor of the seed theory. It can be written as follows
\eq{
W_4(x) 
 = \sum_{i=1}^N  \Big[ (T^{(i)} T^{(i)})(x) - \frac{3}{10} \partial^2 T^{(i)}(x) \Big] -  \frac{\frac{22}{5c} + 1}{N- 1}\sum_{i\ne j} (T^{(i)} T^{(j)})(x)\,. 
}
It is not hard to check that $W_4$ is a Virasoro primary under the Virasoro algebra of Sym$^N({\cal C})$, and thus it is a current of spin four that appears in any symmetric product orbifold with $N>1$. Similarly, permutation invariant combinations of seed stress tensors can be found that make up an infinite tower of higher spin currents (of even spin) as $N$ grows large.
In fact, it has been argued that symmetric product orbifolds have an enlarged symmetry algebra reminiscent of a $\mathcal{W}_N$ higher spin algebra \cite{Baggio:2015jxa,Gaberdiel:2015uca,Gaberdiel:2015mra}. Recently it has been conjectured that all higher spin currents present due to the orbifold structure obtain anomalous dimensions under deformations by single-trace twisted exactly marginal deformations, independent of the details of the seed theory (as long as a twisted marginal deformation exists) \cite{Apolo:2022fya,Benjamin:2022jin}.

On a related note, the orbifold structure also gives rise to an array of non-invertible symmetries. The associated universal defects realize a Rep$(S_N)$ fusion algebra and have been explicitly constructed in \cite{Gutperle:2024vyp}, see also \cite{Knighton:2024noc}. Moreover, the authors show that there exist universal defects that remain nontrivial in the $N\rightarrow\infty$ limit.

These enlarged symmetry algebras present in symmetric product orbifolds are problematic for a holographic interpretation of a local low-energy gravitational description. The presence of towers of higher spin currents prevents the large gap condition of \cite{Heemskerk:2009pn} from being satisfied. The large gap condition can be related to bulk locality. Higher derivative couplings appearing in the bulk EFT can be bounded by inverse powers of the gap \cite{Afkhami-Jeddi:2016ntf, Belin:2019mnx, Kologlu:2019bco, Caron-Huot:2021enk}. Hence, symmetric product orbifolds cannot be dual to a local EFT. 

\paragraph{Hagedorn growth of light states.}

Another universal property that indicates a non-local dual description is the universal growth of the light spectrum of symmetric product orbifolds. As we mentioned above, symmetric product orbifolds not only satisfy the HKS sparseness condition \eqref{eq:HKS}, but they saturate it. More concretely, for any symmetric product orbifold theory at large $N$, the density of states satisfies \cite{Keller:2011xi,Belin:2014fna}
\eq{\label{eq:symhagedorn}
\rho(\Delta)\sim e^{2\pi\Delta}\,,\qquad \text{for}\qquad 1\ll\Delta<\frac{c}{12}\,.
}
This can be recognized as a Hagedorn spectrum of light states, and is interpreted as a stringy spectrum; there are non-local light degrees of freedom associated with string-like states. The string length is of order one in AdS units, meaning that there are non-local effects at the AdS scale in the bulk. This tells us that the bulk dual of symmetric product orbifolds is not Einstein gravity (or supergravity), but it could be a highly curved string theory. 

The growth behavior \eqref{eq:symhagedorn} is sourced by the twisted sectors with long cycles, as we review in App.\,\ref{app:longcycles}. In instances where symmetric product orbifolds appear in the context of string theory, such as in the D1-D5 system, such long-cycle twisted states are dual to long strings \cite{Eberhardt:2018ouy,Eberhardt:2019ywk}. In that context, the holographic dual of the symmetric product orbifold consists of tensionless string theory, which corresponds to the largest possible string length, precisely at the AdS scale.

It is worth commenting on the connection between this Hagedorn growth and the higher spin currents described above. To the best of our knowledge, these are two separate effects, both of which obstruct a standard holographic dual. The density of states sourced by the currents is much slower than Hagedorn and grows like \cite{Gaberdiel:2010ar,Castro:2010ce}
\be
\rho^{\text{HS currents}}(\Delta) \sim e^{\Delta^{2/3} } \,.
\ee
When a marginal deformation exists, like in the D1-D5 CFT, the deformation lifts both the currents and all the string states \cite{Avery:2010er,Gaberdiel:2015uca,Keller:2019yrr,Keller:2019suk,Guo:2020gxm,Benjamin:2021zkn,Apolo:2022fya,Benjamin:2022jin}.

\paragraph{Chaos and OTOC.}

The final universal property we review pertains to chaos. A diagnostic for (quantum) chaos is the out-of-time-ordered correlator (OTOC)
\eq{
\langle \bfO_1(0)\bfO_2(t)\bfO_1(0)\bfO_2(t)\rangle_\beta\,,
}
properly normalized. In a chaotic theory, the OTOC should decay at late times for simple operators. The OTOC is an important probe of an emergent spacetime, in particular the structure of horizons. At late times, the OTOC describes a scattering process that happens very close to what is meant to be a black-hole (or more generally a Rindler-like) horizon. It is really the structure of the horizon that is responsible for the decay of the OTOC, which makes it particularly interesting to study.

It was shown that in all symmetric product orbifolds, the OTOC does not decay at late times \cite{Perlmutter:2016pkf,Belin:2017jli}, leading to the universal conclusion that symmetric product orbifolds are not chaotic. This is quite natural since symmetric product orbifolds should be thought of as analogs of free gauge theories. Their free nature is what produces all the higher-spin currents, and the higher-spin currents are of crucial importance here, since they conspire to prevent the decay of the OTOC that the stress tensor would cause if it was the only current around. Gravity on the other hand is maximally chaotic, as black holes are the fastest scrambling objects known \cite{Sekino:2008he,Lashkari:2011yi,Maldacena:2015waa}. Hence the late-time behavior of the OTOC forms another instance of universality that differs from the one expected for standard holographic interpretation.
\\

Symmetric product orbifolds thus have a large array of universal features. Some of which support a gravitational interpretation in terms of an emergent spacetime, while others obstruct it.\footnote{We have not made an exhaustive list of either: many more properties have been studied (for example boundary states \cite{Belin:2021nck,Gaberdiel:2021kkp}). We have focused here on the most well-understood properties and the ones that are directly related to CFTs being holographic.} The commonly accepted picture is the following: symmetric product orbifolds form a large class of theories that are under good mathematical control and mimic gravity in nontrivial ways, thus forming a rich and interesting playground to study in holography. However, symmetric product orbifolds do not have a gravitational interpretation in terms of an emergent spacetime, possibly weakly coupled to matter, following the laws of general relativity. Rather, they resemble more tensionless string theories in AdS$_3$; see for example \cite{Hikida:2023jyc,Knighton:2024pqh} for a recent discussion for general seed theories.

It should be noted though that at least in certain specific examples, i.e., for some particular choices of the seed theory, symmetric product orbifolds do have a connection to emergent local theories of gravity. In those setups, the symmetric product orbifold contains exactly marginal deformations that are interpreted as gauge couplings that can break the orbifold structure and drive the theory to a holographic point. The best understood example of this is when the seed theory is a nonlinear sigma model on $\mathbb{T}^4$ or K3, which appear in the aforementioned D1-D5 system. But more recently, a whole landscape of symmetric product orbifolds and their deformations have been studied in connection with a holographic point on their conformal manifold, see e.g. \cite{Belin:2020nmp,Apolo:2022fya,Benjamin:2022jin,Apolo:2022pbq,Apolo:2024hoa}. The general picture is that while symmetric product orbifolds display some of the properties of gravity, the other unwanted properties get wiped out as the coupling is increased.

\section{Universality in thermal correlators \label{sec:thermaluniv} }

In the remainder of this work, we will demonstrate other universal aspects of symmetric products encoded in its correlation functions, and discuss the implications of our findings for a holographic interpretation in terms of an emergent spacetime. This section focuses on thermal one- and two-point functions of the theory. These correlation functions at finite temperature are prominent when discussing the physics or black holes, in particular, the connection between the spectral density of the thermal two-point function and the emergence of spacetime through the use of Von Neumann algebras, which we will address in Sec.\,\ref{sec:grav}.  

The strategy we use to study thermal correlators is first to understand how to compute correlation functions on the torus, discussed in Sec.\,\ref{sec:torus}, after which we restrict the modular parameter of the torus to $\tau=i\beta$, giving rise to the correlation function at finite temperature. As we will see in Sec.\,\ref{sec:thuni}, universality of the thermal correlators emerges once the large-$N$ limit is taken. The origin of this universality coincides with that of the partition function mentioned in \eqref{eq:symhagedorn}.

\subsection{Correlation functions on the torus}\label{sec:torus}

Compared to correlation functions on the plane, correlation functions on more complicated manifolds can be intractable or tedious at best. Recently, correlation functions of symmetric product orbifold theories on generic manifolds were considered in \cite{Aharony:2024fid}. There, the authors derived a generating function for the partition function with sources on a generic manifold in the grand canonical ensemble. This generating function exponentiates into a sum over connected components of all possible cover surfaces of the manifold. Because our main interest is in torus correlation functions we can be much more explicit. In fact, the torus two-point function was derived explicitly in the grand canonical ensemble for single-trace untwisted operators by \cite{Kyriakosunpublished}. From the generating functions, one can formally extract the fixed $N$ correlators. Here, we will find an explicit formula at fixed $N$ that is convenient to work with. Our formalism follows that of Bantay, which has the advantage of being valid for any permutation orbifold. 

The starting point is Bantay's formula for the torus partition function \eqref{eq:bantay}. We would like to extend \eqref{eq:bantay} to allow for operator insertions on the torus. This is feasible in the case where the operators inserted are untwisted. Including twisted insertions is harder because in that case, the cover surfaces generated are of higher genus.\footnote{This is related to the trade-off made in \cite{Aharony:2024fid}. The generating function there is also valid for twisted sector insertions. A consequence of this generality is that the expression is in terms of a sum over (implicit) cover surfaces.} For simplicity we will restrict ourselves to single-trace untwisted operators, but it should be straightforward to generalize to untwisted multi-trace insertions by carefully keeping track of the combinatorics involved in multi-trace operators. We draw the reader to an important comment on notation: in this section we will not include the standard normalization by the torus partition function. More precisely, we will write\footnote{To simplify notation, we will suppress the anti-holomorphic dependence when possible in the rest of this section.}
\eq{
\left\langle\bfO_1(z_1,\bar{z}_1)\cdots\bfO_n(z_n,\bar{z}_n)\right\rangle_\tau\coloneqq \Tr\left[q^{L_0-c/24}\bar{q}^{\bar{L}_0-c/24}\bfO_1(z_1,\bar{z}_1)\cdots\bfO_n(z_n,\bar{z}_n)\right]\,,
}
where $q\coloneqq e^{2\pi i\tau}$. The conventional normalization for thermal correlators would then be expressed as our unnormalized correlators divided by the partition function
\be
\frac{\left\langle\bfO_1(z_1,\bar{z}_1)\cdots\bfO_n(z_n,\bar{z}_n)\right\rangle_\tau}{Z(\tau,\bar{\tau})} \,.
\ee

The untwisted single-trace operators we consider can always be written as
\eq{\label{eq:sto}
\bfO(z)=\frac{1}{\sqrt{N}}\sum_{j=1}^N \bfOs^{(j)}(z)\,.
}
To insert $\bfO$ on the torus, all we have to do is add the seed insertions $\bfOs$ appropriately on the cover. Since the torus has two periodic directions, we have to include images of the seed operator along both cycles on the cover. In order to limit the notation, we will introduce a function $G$, that depends on the modular parameter $\tau$, seed insertions $\bfO_{\text{seed}}(z_i)$, and an orbit $\xi$. With this data, $G$ provides the cover surface associated with the orbit and the insertions corresponding to the seed operators with the appropriate periodic images on the cover. Concretely, $G$ is given by\footnote{Note that in principle the seed insertions do not have to be of the same operator. If we include operators of different conformal dimensions, the only change to this formula is that in the power of $\lambda_\xi$ we should sum over the conformal weight of the different insertions. Moreover, as mentioned before, we have suppressed the anti-holomorphic dependence on $\bar{\tau}$ and $\bar{z}_i$.}
\eqsp{
G(\xi,\tau,\bfO_{\text{seed}}(z_1),\cdots\bfO_{\text{seed}}(z_n))&\coloneqq\sum_{j_1,\dots,j_n=0}^{\lambda_\xi-1}\sum_{\ell_1,\dots,\ell_n=0}^{\mu_\xi-1}\lambda_\xi^{-n \Delta_{\bfO}}\\
&\hspace{-60pt}\left\langle\bfO_{\text{seed}}\left(\frac{z_1+ 2\pi j_1+ 2\pi \ell_1\tau}{\lambda_{\xi}}\right)\cdots \bfO_{\text{seed}}\left(\frac{z_n+ 2\pi j_n+2\pi \ell_n\tau}{\lambda_{\xi}}\right)\right\rangle_{\tau_{\xi}}\,,
}
where $\Delta_{\bfO}$ is the conformal dimension of $\bfO$. We have inserted $\bfOs$ on a (connected component of a) covering torus, which instead of parameter $\tau$ now has modular parameter $\tau_{\xi}$ \eqref{eq:tauxi}, repeated here for convenience
\eq{\label{eq:tauxi2}
\tau_\xi=\frac{\mu_\xi\tau+\kappa_\xi}{\lambda_\xi}\,.
} 
Here $\kappa_\xi$ is non-negative while $\lambda_\xi$ and $\mu_\xi$ are positive integers and depend on the details of the orbit $\xi$. The cover torus with modular parameter $\tau_\xi$ is bigger than the original one; the cycles have increased by a factor $\lambda_{\xi}$ and $\mu_{\xi}$ respectively. Under this rescaling of the torus, also the insertion point $z$ changes. Under a modular transformation $\tau\mapsto\frac{a\tau+b}{c\tau+d}$, $z$ gets mapped to $\frac{z}{c\tau+d}$. In our case $a=\mu_{\xi}$, $b=\kappa_{\xi}$, $c=0$, and $d=\lambda_{\xi}$. Due to this conformal transformation, we have to include the Jacobian transformation of the primary operator $\bfOs$ given by $\lambda_{\xi}^{-\Delta_\bfO}$. Finally, we have to include the nontrivial images of $z$ on the enlarged torus, which is achieved by the sums over $j$ and $\ell$.

Now that we understand how to insert seed operators on the cover tori with appropriate periodicities, we can turn to the evaluation of torus correlation functions in permutation orbifolds. We will give explicit expressions for the torus one- and two-point functions; adding more insertions is a straightforward generalization of these cases.

To get the torus one-point function in a permutation orbifold we now just have to sum over all possible coverings and include the operator insertions appearing in \eqref{eq:sto} appropriately. The result is
\eqsp{\label{eq:1pt}
&\left\langle\bfO(z)\right\rangle_\tau=\frac{1}{\sqrt N\abs{G_N}}\sum_{\substack{g,h\in G_N\\gh=hg}}\sum_{\xi\in O(g,h)}G(\xi,\tau,\bfOs(z))\prod_{\substack{\xi'\in O(g,h)\\ \xi'\neq\xi}}Z_\mathcal{C}(\tau_{\xi'})\,.
}
This equation can be understood as follows. We sum over the seed insertions on connected components of all possible cover surfaces. The cover surface, as explained in Sec.\,\ref{sec:z}, is picked out by a pair of commuting elements $g$ and $h$, and the connected components of such a covering are labeled by the orbits $\xi\in O(g,h)$. We sum over the insertions (with the appropriate periodic images) on each of those connected components, while for the other components, denoted here by $\xi'$, we include the appropriate seed partition function.

Next, we consider the torus two-point function. The structure of the torus two-point function for permutation orbifolds is similar to the one-point function. We sum over all possible seed insertions on the various cover surfaces of the torus. The resulting expression is given by
\eqsp{
\left\langle\bfO(z_1)\bfO(z_2)\right\rangle_\tau=\frac{1}{N\abs{G_N}}\sum_{\substack{g,h\in G_N\\gh=hg}}\Bigg[\sum_{\xi\in O(g,h)}G(\xi,\tau,\bfOs(z_1),\bfOs(z_2))\prod_{\substack{\xi'\in O(g,h)\\ \xi'\neq\xi}}Z_\mathcal{C}(\tau_{\xi'})\\
+\sum_{\substack{\xi_1,\xi_2\in O(g,h)\\ \xi_1\neq\xi_2}}G(\xi_1,\tau,\bfOs(z_1))G(\xi_2,\tau,\bfOs(z_2))\prod_{\substack{\xi'\in O(g,h)\\ \xi_1\neq\xi'\neq\xi_2}}Z_\mathcal{C}(\tau_{\xi'})\Bigg]\,.\label{eq:2pt}
}
There are two qualitatively different contributions. The contribution on the first line originates from the two seed insertions being inserted on the same connected component of the cover surface, while on the second line the insertions are on different connected components. Note that the two lines of \eqref{eq:2pt} are not precisely equal to the connected and disconnected part of the two-point function. The connected thermal two-point function is
\eqsp{
\left\langle\bfO(z_1)\bfO(z_2)\right\rangle_\tau^{\text{connected}}&=\left\langle\bfO(z_1)\bfO(z_2)\right\rangle_\tau-\left\langle\bfO(z_1)\right\rangle_\tau\left\langle\bfO(z_2)\right\rangle_\tau\\
&=\frac{1}{N\abs{G_N}}\sum_{\substack{g,h\in G_N\\gh=hg}}\sum_{\xi\in O(g,h)}\Bigg[G(\xi,\tau,\bfOs(z_1),\bfOs(z_2))\\
&\hspace{40pt}-G(\xi,\tau,\bfOs(z_1))G(\xi,\tau,\bfOs(z_2))\Bigg]\prod_{\substack{\xi'\in O(g,h)\\ \xi'\neq\xi}}Z_\mathcal{C}(\tau_{\xi'})\,.
}

It is not hard to see that \eqref{eq:2pt} has the expected coincidence limit. The only ingredient we need to show this is the coincidence limit for the (normalized) torus two-point function in a generic CFT $\mathcal{C}$. For simplicity, we quote the result for scalars with $h=\bar{h}=\frac{\Delta}{2}$, but it is straightforward to generalize to other spins. We have
\eq{
\frac{\left\langle\bfO(z_1,\bar{z_1})\bfO(z_2,\bar{z}_2)\right\rangle_\tau}{Z_{\cal C}(\tau)}\underset{z_1\rightarrow z_2}{=}\frac{1}{|z_1-z_2|^{2\Delta_\bfO}}+\cdots\,,
}
with $Z_{\cal C}(\tau)$ the torus partition function of $\mathcal{C}$. Applying this fact to the right-hand side of \eqref{eq:2pt} we find
\begin{align} \notag\frac{\left\langle\bfO(z_1,\bar{z_1})\bfO(z_2,\bar{z}_2)\right\rangle_\tau}{Z_{G_N}(\tau)}&\underset{z_1\rightarrow z_2}{=}\frac{1}{Z_{G_N}(\tau)}\frac{1}{N\abs{G_N}}\sum_{\substack{g,h\in G_N\\gh=hg}}\sum_{\xi\in O(g,h)}\frac{\abs{\xi}}{|z_1-z_2|^{2\Delta_\bfO}}Z(\tau_{\xi})\prod_{\substack{\xi'\in O(g,h)\\ \xi'\neq\xi}}Z_{\mathcal{C}}(\tau_{\xi'})+\cdots\\
    &=\frac{1}{Z_{G_N}(\tau)}\frac{1}{|z_1-z_2|^{2\Delta_\bfO}}\frac{1}{\abs{G_N}}\sum_{\substack{g,h\in G_N\\gh=hg}}\prod_{\xi\in O(g,h)}Z_{\mathcal{C}}(\tau_\xi)+\cdots\\
    &=\frac{1}{|z_1-z_2|^{2\Delta_\bfO}}+\cdots\,.\notag
\end{align}
Here we used in the first line that there is a divergence in the coincidence limit on the right-hand side of \eqref{eq:2pt} whenever $j_1=j_2$ and $\ell_1=\ell_2$, and there are precisely $\lambda_\xi\mu_\xi=\abs{\xi}$ instances of this occurrence. Furthermore, the sum over orbits $\xi$ leads to a factor of $N$ once the lengths of the orbits $\abs{\xi}$ are added up. Finally, in the last line we used \eqref{eq:bantay}. We can thus conclude that \eqref{eq:2pt} has the expected coincidence limit.

Using the same formalism and techniques we used to derive \eqref{eq:2pt} it is now a straightforward task to find expressions for higher-point functions of untwisted single-trace operators on the torus. Closely related to higher-point functions of single-trace operators is also the insertion of multi-trace operators, which can be found by taking coincidence limits of (some of the) single-trace operator insertions. We will not analyze such higher-point and multi-trace correlators here, but instead shift our focus to the universality at large $N$ of the correlator \eqref{eq:2pt}.

\subsection[Universality in the thermal limit at large \texorpdfstring{$N$}{N}]{Universality in the thermal limit at large $\boldsymbol{N}$}\label{sec:thuni}

Now that we have expressions for the torus one- and two-point functions, we turn to the question of universality in the large-$N$ limit. We will mostly focus on the high-temperature regime $\beta<2\pi$. The behavior at low temperatures can be inferred via modular invariance, and we will comment on low temperatures at the end of this section. From the analysis of \cite{Keller:2011xi,Belin:2014fna}, reviewed in App.\,\ref{app:longcycles}, it follows that, for high temperatures $\beta<2\pi$, the leading contributions to the sum over cover surfaces come from long twist cycles. In other words, instances in the sum over $g,h$ where $g$ has at least one long cycle. For simplicity, we restrict to $h=(1)$, and $g=(1)^{N-L}(L)$.\footnote{Note that for more general permutation groups than $S_N$, this permutation $g$ is not necessarily an element of the group. It is not hard to see, however, that as long as $g$ has a long cycle in it, the qualitative picture we sketch below persists.} This cover surface has $N-L+1$ connected components, all of which have $\mu_\xi=1$ and $\kappa_\xi=0$. There is one connected component with $\lambda_\xi=L$ (coming from the cycle of length $L$), while the other $N-L$ connected components have $\lambda_\xi=1$.

We start with the large-$N$ behavior of the one-point function \eqref{eq:1pt}. There are two qualitatively different contributions to the one-point function coming from the cover surface discussed above: one with the operator inserted on the $(L)$-cycle, and $N-L$ contributions with the operator inserted on one of the trivial cycles. These contributions schematically look like
\eq{
\frac{N-L}{\sqrt{N}}\left\langle\bfOs(z)\right\rangle_\tau Z_\mathcal{C}(\tau)^{N-L-1}Z_\mathcal{C}\left(\frac{\tau}{L}\right)+
\frac{L}{\sqrt{N}}\left\langle\bfOs\left(\frac{z}{L}\right)\right\rangle_{\tau/L} Z_\mathcal{C}(\tau)^{N-L}\,.\label{eq:l1pt}
}
Here in the last term we have not denoted all periodic images explicitly but included a factor $L$ out front because there are $L$ terms coming from the different images, and the one-point on the torus is translationally invariant. One might worry that \eqref{eq:l1pt} diverges as we take $L$ to scale with $N\gg1$. Naively, the prefactor of the second term in \eqref{eq:l1pt} indeed scales like $\sqrt{N}$ as $L\sim N$. However, the effective temperature of that term is $\tau_{\text{eff}}=\frac{i\beta}{2\pi L}$ which goes to infinite temperature at large $L$, and thus the one-point function multiplying this large prefactor is exponentially suppressed in $L$ \cite{Kraus:2016nwo}, i.e.\footnote{Note that in this equation, the one-point function is always exponentially suppressed irrespective of the scaling dimension of the seed operator. To see this, we need to add in the proper normalization by the partition function, which cancels the central charge dependence in the exponent. The resulting expression for the properly normalized correlator scales like $e^{-4\pi^2L\Delta_\chi/\beta}$.}
\eq{
\left\langle\bfOs\left(\frac{z}{L}\right)\right\rangle_{\frac{i\beta}{2\pi L}}=i^{J_\bfOs}C_{\chi\chi\bfOs}\left(\frac{2\pi L}{\beta}\right)^{\Delta_\bfOs}e^{-4\pi^2L(\Delta_\chi-c/12)/\beta}+\cdots\,, 
}
with $\chi$ the lowest weight seed operator such that $C_{\chi\chi\bfOs}\neq 0$, and corrections to this leading order term are further exponentially suppressed in $L$. Note that we only discussed one particular cover that contributes to the one-point function. It is not hard to see that the behavior remains the same as we consider more general cover surfaces, as long as there is a long twist cycle involved which scales in the large-$N$ limit.

Next, we turn to the two-point function \eqref{eq:2pt}, and study the leading order behavior of thermal two-point functions in permutation orbifolds at large $N$. We focus on the contribution coming from the first line of \eqref{eq:2pt}. The ``disconnected" contribution on the second line is suppressed in $N$ by the same arguments we used for the thermal one-point function. As is the case for the thermal one-point function, there are two qualitatively different contributions given by
\begin{align} \label{eq:approx2}
\begin{split}
\frac{1}{N}\Bigg(\sum_{j_1,j_2=0}^{L-1}L^{-2h_\bfO}\left\langle\bfOs\left(\frac{z_1+2\pi j_1}{L}\right)\bfOs\left(\frac{z_2+2\pi j_2}{L}\right)\right\rangle_{\tau/L} Z_\mathcal{C}(\tau)^{N-L}\\
+(N-L)\left\langle\bfOs(z_1)\bfOs(z_2)\right\rangle_\tau Z_\mathcal{C}(\tau)^{N-L-1}Z_\mathcal{C}(\tau/L)\Bigg)\,.
\end{split}
\end{align}
Here the contribution on the first line comes from inserting the seed insertions on the cycle of length $L$, while on the second line, the seed operators are inserted in one of the trivial cycles. The thermal two-point function is now found by setting $\tau=i\beta$. We use that in any CFT, in particular in the seed CFT, the thermal two-point function at high temperatures $\beta\rightarrow0$ is given by
\eq{\label{eq:3.13}
\left\langle \bfO(z_1)\bfO(z_2)\right\rangle_\beta\underset{\beta\rightarrow 0}{\sim}\left(\frac{\beta}{\pi}\right)^{-2h_\bfO}\frac{1}{\sinh\left(\frac{\pi(z_1-z_2)}{\beta}\right)^{2h_\bfO}}\,.
}
In the limit where $L\sim N\gg1$ we then obtain for the first term of \eqref{eq:approx2}
\eqsp{
&\frac{1}{N}\sum_{j_1,j_2=0}^{L-1}L^{-2h_\bfO} \left(\frac{\pi L}{\beta}\right)^{2h_\bfO}\frac{1}{\sinh\left(\frac{\pi(z_1-z_2+2\pi(j_1-j_2))}{\beta}\right)^{2h_\bfO}}
\\
&\hspace{180pt}\sim \sum_{j\in\mathbb{Z}}\frac{1}{\left(\frac{\beta}{\pi}\sinh\left(\frac{\pi(z_1-z_2+2\pi j)}{\beta}\right)\right)^{2h_\bfO}}\,.\label{eq:th2pt}
} 
In the first line, we used \eqref{eq:3.13}, which is valid when $L$ is large, and in the second line, we exchanged one of the sums for a factor of $L\sim N$, which is possible because the terms in the sum only depend on $j_1-j_2$.
Here it is essential that the temperature at which we evaluated the seed thermal two-point function is enhanced by a factor proportional to $L$. This implies that even at finite $\beta$ in the orbifold theory, we can use the universal high-temperature answer for the seed thermal two-point functions used to compute the orbifold two-point function. Furthermore, the second line of \eqref{eq:approx2} is suppressed by the prefactor $\frac{N-L}{N}$ as $L\sim N$.

The long twist cycle has two crucial effects for a holographic interpretation. First, it gives rise to the sum over images, which comes from the periodic images of the seed insertion on the cover surface, and second it increases the temperature at which we evaluate the seed two-point functions by a factor proportional to $N$. Hence the final result is an expression that includes the sum over images and has an extended regime of validity compared to the correlation function at finite temperature for a generic CFT$_2$.

\subsection*{Regimes of validity}

We emphasize once more two important remarks. First, not every permutation orbifold contains the specific permutations we considered here, i.e., $g=(1)^{N-L}(L)$. Symmetric product orbifolds however do contain this permutation, and form our main interest. Second, we only considered one very specific permutation, whereas \eqref{eq:2pt} instructs us to sum over all pairs of commuting elements $g$ and $h$. Intuition from the thermal partition function tells us that the main role of the different choices of $h$ is to symmetrize over the full permutation group. Moreover, it has been argued that the leading contribution to the sum over commuting elements of $S_N$ comes from long twist cycles, that is, choices for $g$ with long cycles. It is not hard to see that for any choice of $g$ with a long cycle, our qualitative conclusions remain unchanged; we still generate a sum over images, and the temperature at which we evaluate the seed two-point function is enhanced by the length of the cycle. Hence we can argue that \eqref{eq:th2pt} is the universal leading order behavior of the thermal two-point function of untwisted single-trace operators in any symmetric product orbifold in the large-$N$ limit. However, a more careful discussion of the regime of validity of \eqref{eq:th2pt} is warranted.

The argument we use to understand the regime of validity of \eqref{eq:th2pt} is as follows. Since we consider light insertions on a high-temperature object (the partition function), the insertion of these light operators influences order one dynamics, but they are not expected to change the large-$N$ considerations. Hence, we expect the regime of validity not to depend on the details of the insertions (at least at leading order), but rather to be determined by the large-$N$ behavior of the torus partition function. This is a general effect of saddle contributions to path integrals, where probes of saddles typically do not affect which saddle dominates, at least if we stay sufficiently far from phase transitions. For generic permutation orbifolds, the phase structure of the partition function is not completely pinned down. It has been shown that there is always a regime of Hagedorn growth \cite{Belin:2014fna}, but for generic choices of $G_N\subseteq S_N$ the HKS sparseness condition is not satisfied, and hence the phase structure is unclear. In the case of symmetric product orbifolds, $G_N=S_N$, we can say a lot more. 

As discussed in Sec.\,\ref{sec:symuniversality}, the density of states of symmetric product orbifolds universally exhibits Hagedorn growth at temperatures $\beta>2\pi$ and has a Hawking-Page phase transition at $\beta=2\pi$, above which the growth is characterized by Cardy behavior \cite{Keller:2011xi,Belin:2014fna}. Thus, in symmetric product orbifolds the universal Cardy growth at high temperatures $\beta\ll1$ has an extended regime of validity all the way down to the phase transition at $\beta=2\pi$. The states in symmetric product orbifolds responsible for this universality are precisely the twisted sectors with long cycles we considered when deriving \eqref{eq:th2pt}. As a result, we expect the regime of validity of the universal behavior of the thermal two-point function to extend all the way down to the Hawking-Page phase transition. The conclusion of our considerations is thus that above the Hawking-Page transition, the thermal two-point function takes the universal form
\eq{\label{eq:finalth}
\left\langle\bfO(z_1)\bfO(z_2)\right\rangle_\beta=\sum_{j\in\mathbb{Z}}\frac{1}{\left(\frac{\beta}{\pi}\sinh\left(\frac{\pi(z_1-z_2+2\pi j)}{\beta}\right)\right)^{2h_\bfO}}+\cdots\,,\qquad \text{for}\qquad \beta<2\pi\,.
}
Here, the $\cdots$ mean we only write the leading term in the large-$N$ limit. We neglect them below, and one should understand our expressions as the leading large $N$ terms. In order to facilitate the gravitational interpretation of this result, is useful to write it in terms of the coordinates $z=\phi+it_{\rm E}$. Here $\phi$ is interpreted as an angular coordinate, periodic with period $2\pi$, whereas $t_{\rm E}$ should be interpreted as Euclidean time with period $\beta$. Furthermore, we restore the anti-holomorphic dependence and remind the reader that for simplicity we are considering scalar probes, \ie, $h_\bfO=\bar{h}_\bfO$. We find\footnote{Note that Maldacena already found this expression for a particular thermal two-point function in the context of the D1-D5 CFT \cite{Maldacena:2001kr}.}
\eqsp{\label{eq:tphires}
\left\langle\bfO(t_{\rm E},\phi)\bfO(0)\right\rangle_\beta=\sum_{j\in\mathbb{Z}}\frac{1}{\left(\frac{\beta}{\pi}\sinh\left(\frac{\pi(\phi+it_{\rm E}+2\pi j)}{\beta}\right)\right)^{2h_\bfO}\left(\frac{\beta}{\pi}\sinh\left(\frac{\pi(\phi-it_{\rm E}+2\pi j)}{\beta}\right)\right)^{2h_\bfO}}\\
=\sum_{j\in\mathbb{Z}}\frac{1}{\left(\frac{\beta^2}{4\pi^2}\left(\cosh\left(\frac{2\pi(\phi+2\pi j)}{\beta}\right)-\cos\left(\frac{2\pi t_{\rm E}}{\beta}\right)\right)\right)^{2h_\bfO}}\,,
}
where, as before, this expression is valid at large $N$ above the phase transition $\beta<2\pi$. We can now easily identify also the low-temperature behavior via a modular $S$-transformation. In terms of the coordinates the $S$-transformation just exchanges the angular coordinate and Euclidean time $\phi\leftrightarrow t_{\rm E}$. Furthermore, the inverse temperature is reflected around its self-dual value of $2\pi$, that is, the $S$-transformation sends $\beta\to4\pi^2/\beta$. Hence we end up with the following expression for the thermal two-point function
\eq{\label{eq:lowt}
\hspace{-10pt}\left\langle\bfO(t_{\rm E},\phi)\bfO(0)\right\rangle_\beta=\begin{cases}
    \displaystyle\sum_{j\in\mathbb{Z}}\frac{1}{\left(\frac{\beta^2}{4\pi^2}\left(\cosh\left(\frac{2\pi(\phi+2\pi j)}{\beta}\right)-\cos\left(\frac{2\pi t_{\rm E}}{\beta}\right)\right)\right)^{2h_\bfO}}\quad&\text{for}\quad\beta<2\pi\\
    \displaystyle\sum_{j\in\mathbb{Z}}\frac{1}{\left(\frac{4\pi^2}{\beta^2}\left(\cosh\left(\frac{\beta(t_{\rm E}+2\pi j)}{2\pi}\right)-\cos\left(\frac{\beta \phi}{2\pi}\right)\right)\right)^{2h_\bfO}}\quad&\text{for}\quad\beta>2\pi
\end{cases}\,.
}
It is interesting to comment on the terms in the sum over commuting elements $g,h$ in \eqref{eq:2pt} that dominate at low temperatures. Again, we can utilize the $S$-transformation. At high temperatures, the orbits leading to $\tau_\xi=\tau/L$ were crucial for the leading behavior of the thermal correlators (and partition function). In the low-temperature regime, the dominant contribution thus comes from terms like $\tau_\xi= \tau L$. Such terms are sourced by taking $g=(1)$ and $h=(L)$ in the sum \eqref{eq:2pt}. Hence, the low-energy behavior is dominated by the untwisted sector. In fact, the sum over images that appears in the low-temperature expression can be tied to the contribution of multi-trace operators in the Boltzmann sum, see App.\,\ref{app:hks2pt}.

The results of this section form an interesting explicit realization of the analysis of \cite{Kraus:2017kyl}. There, under the assumption of a sparse light spectrum, large-$N$ factorization, and some mild additional assumptions on OPE coefficients, it is shown that the thermal two-point function undergoes a phase transition at $\beta=2\pi$, and in either phase can be written as a sum over light OPE coefficients. The interpretation given is that above the Hawking-Page transition, the thermal two-point function is given by a sum over Witten diagrams in a BTZ geometry. In App.\,\ref{app:hks2pt} we present a summary of the results in \cite{Kraus:2017kyl} and show that symmetric product orbifolds satisfy the required constraints.

\section{Universality in correlation functions of twisted ground states \label{sec:twisteduniv}}

In this section, we will compute two-point functions in particular heavy states of symmetric product orbifolds, and study their universality. Due to the state-operator correspondence, these heavy-state two-point functions are four-point functions. A four-point correlator generically depends on OPE coefficients and hence on dynamical aspects of the CFT. For this reason, universality in four-point functions is not expected. Symmetric product orbifolds, however, contain universal sectors of operators: the twisted sectors. In particular, the twisted sector ground states have certain universal properties: their dimension and some of their OPE coefficients depend only on the central charge of the seed theory \cite{Burrington:2018upk}, and on none of its other details. There may thus be some universality in four-point functions containing twisted sector insertions, in particular at large $N$. Investigating this question is the topic of this section.

In the context of four-point functions in symmetric product orbifolds, a correlation function will be coined universal if it does not depend on seed OPE coefficients. For this reason, we will focus on four-point functions that contain at least two insertions of twisted ground states. That is, we consider (connected) four-point functions on the cylinder of the form\footnote{Note that the twisted ground state operators do not necessarily have to be identical to obtain a nonzero answer for the correlator. Nevertheless, we will consider the twisted ground state insertions to be identical.}
\begin{equation}\label{eq:4ptgen}
    \langle \sigma_{[g]}(w_1) \bfO (w_2) \bfO (w_3) \sigma_{[g]}(w_4)\rangle_{\rm c}\,,
\end{equation}
where $\bfO$ is a single-trace operator in Sym$^N({\cal C})$ (for now we do not specify any further details of $\bfO$), $\sigma_{[g]}$ is the ground state of the twist-$[g]$ sector, and we restrict to the connected component. We emphasize that in this section, we use the term connected differently than usual. With the subscript $\rm c$ we will mean connected in the sense of the copies of the symmetric product orbifold, but keep the entire seed correlators (also the disconnected piece). For example, for untwisted single-trace operators we will denote 
\[\langle \bfO(z_1)\bfO(z_2)\bfO(z_3)\bfO(z_4)\rangle_{\rm{c}}=N^{-2}\sum_i\langle \bfO^{(i)}(z_1)\bfO^{(i)}(z_2)\bfO^{(i)}(z_3)\bfO^{(i)}(z_4)\rangle\,.
\]
In contrast, a disconnected term has seed insertions on at least two distinct copies. This means that a disconnected term with our convention is always disconnected in the usual sense.

As mentioned, the operators $\bfO$ included in \eqref{eq:4ptgen} are single trace. They can for example be untwisted and be generated by inserting an operator of the seed CFT, $\bfOs$, on the $i$-th copy of the symmetric product, which we denote by $\bfO^{(i)}$. Alternatively, they can be twisted, and we will denote an operator in the twist-$[g]$ sector with $\bfO_{\sigma_{[g]}}$. The operator will be single-trace if $[g]$ corresponds to a conjugacy class with a single non-trivial cycle. As reviewed in Sec.\,\ref{sec:RevCorr}, a correlation function with twisted sector insertions can be evaluated by going to a cover surface. In the process of going to the cover space the effects of the twists involved are undone. Therefore, on the cover, \eqref{eq:4ptgen} reduces to a two-point function of the seed operators involved in $\bfO$ and one can hope to obtain a universal answer for \eqref{eq:4ptgen}.

Without loss of generality, we can use conformal invariance to map the insertion points $(w_1,w_2,w_3,w_4)$ in \eqref{eq:4ptgen} to $(\infty,0,w,-\infty)$. The correlator can then be regarded as the two-point function of $\bfO$ in the twisted ground state $\ket{\sigma_{[g]}}$. We can thus say that $\bfO$ is an operator that probes the state created by $\sigma_{[g]}$. In the remainder of the section we will denote the operator probing the twist ground states with $\bfOp$.

Twisted ground states are fully characterized by partitions of integers, which are in one-to-one correspondence to the conjugacy classes of the symmetric group. We will write integer partitions of $N$ as sequences $\{N_n\}_{n\in\mathbb{N}}$ with
\begin{equation}
N=\sum_{n=1}^\infty nN_n\,.
\end{equation}
The sequence $\{N_n\}_{n\in\mathbb{N}}$ specifies the distribution of single-cycle constituents of the twisted sector ground state of total twist length $N$
\begin{equation}\label{eq:twistgs}
\sigma_{[g]}=\prod_n \left(\sigma_n\right)^{N_n}\,,
\end{equation}
where the partition $\{N_n\}_{n\in\mathbb{N}}$ corresponds to the conjugacy class $[g]$. Note that this expression has to be appropriately normalized on both sides and should be understood as specifying the branch structure of the operator. As explained in Sec.\,\ref{sec:RevCorr}, in order to compute correlation functions of twisted operators, it is convenient to write them as a sum over twist operators of representatives of the conjugacy class $[g]$. Concretely, we can write
\eq{
\sigma_{[g]}=\frac{1}{\sqrt{N!\abs{\text{Cent}[g]}}}\sum_{h\in S_N}\sigma_{hgh^{-1}}\,.
}
Here we have included the appropriate normalization that ensures a unit two-point function. The normalization factor includes the order of the centralizer subgroup of $g$ in $S_N$, which is equal to
\eq{
\text{Cent}[g]=S_{N_1}\times\left(S_{N_2}\rtimes \mathbb{Z}_2^{N_2}\right)\times\left(S_{N_3}\rtimes \mathbb{Z}_3^{N_3}\right)\cdots\,,
}
and thus has order
\eq{
\abs{\text{Cent}[g]}=\prod_n N_n! \ n^{N_n}\,.
}

Depending on the choice of $\bfOp$, one or more single-cycle contributions to $\sigma_{[g]}$ play a role in evaluating the different connected components of \eqref{eq:4ptgen}, as we will showcase in the remainder of this section. We start by considering $\bfOp$ in the untwisted sector in Sec.\,\ref{sec:notwist}. This is the simplest case: only one $n$ at a time will be involved in the computation of \eqref{eq:4ptgen}. Next, we graduate to probes consisting of twist-2 operators in Sec.\,\ref{sec:addtwist}, where the situation is more complicated. There we will see that two $n$'s can contribute at the same time.

\subsection{Untwisted probes}\label{sec:notwist}

Let us start with the simplest case, we pick as probe $\bfOp$ an untwisted single-trace operator, $\bfO$. We will also further assume that $\bfO$ is primary such that it can be constructed by placing a seed primary $\bfOs$ on the $i$-th copy of the symmetric product, which we shall denote by $\bfO^{(i)}$, and then symmetrizing:
\begin{equation}
    {\bfOp} \coloneqq \bfO= \frac{1}{\sqrt{N}}\sum_i {\bf O}^{(i)}.
\end{equation}
Note that with this choice of the probe operator $h_{\text{probe}}=h_{\bfO}$. 

We are interested in the connected contribution to the correlator \eqref{eq:4ptgen}. Because the probe operators are untwisted, the individual cycles making up the twisted ground state $\sigma_{[g]}$ have to pair up between the two twisted insertions. Moreover, only cycles that multiply to the identity lead to a nonvanishing contribution, in accordance with \eqref{eq:id}. Furthermore, only the terms in the sums where the probe insertions are included within the same cycle lead to a nonzero answer. This can be seen by the fact that if the untwisted operators are inserted on different cycles, one ends up with a one-point function on the cover sphere, which vanishes. The only nonzero contributions to \eqref{eq:4ptgen} are thus proportional to 
\eq{\label{eq:singlecycle4pt}
\langle \sigma_n (\infty)\bfO(0)\bfO(w)\sigma_n(-\infty)\rangle_{\rm c}\,.
}
In the remainder of this subsection, we will evaluate \eqref{eq:singlecycle4pt}.

As laid out in Sec.\ref{sec:rev}, we start by writing \eqref{eq:singlecycle4pt} as a sum over all representatives of the conjugacy class of a single cycle of length $n$
\eq{\label{eq:untwsums}
\frac{1}{N}\frac{1}{n (N-n)!N!}\sum_{j_1,j_2=1}^N\sum_{h_1,h_2\in S_N}\left\langle \sigma_{h_1(1\cdots n)h_1^{-1}}(\infty)\bfO^{(j_1)}(0)\bfO^{(j_2)}(w)\sigma_{h_2(n \cdots 1)h_2^{-1}}(-
\infty)\right\rangle_{\rm c}\,.
}
We can simplify this expression by fixing the gauge $h_1=(1)$, and in turn, multiplying by $\abs{S_N}=N!$\,. This forces $h_2$ to be cyclic on the first $n$ copies, since $(1\cdots n)^{-1}=(n\cdots 1)$. There are thus $\abs{S_{N-n}} \cdot\abs{\mathbb{Z}_n}$ terms left to consider in the sum over $h_2$, which all lead to the same answer. Furthermore, only the values $j_1,j_2\in\{1,\cdots,n\}$ give a connected contribution. Combining all of the above, we arrive at
\eq{\label{eq:untwOO}
\hspace{-10pt}\langle \sigma_n (\infty)\bfO(0)\bfO(w)\sigma_n(-\infty)\rangle_{\rm c}=\frac{1}{N}\sum_{j_1,j_2=1}^n\left\langle \sigma_{(1\cdots n)}(\infty)\bfO^{(j_1)}(0)\bfO^{(j_2)}(w)\sigma_{(n \cdots 1)}(-
\infty)\right\rangle\,.
}

To compute the correlator in the sum, we move to a suitable cover space. The map from base to cover space is easiest when the base space is flat, so as an intermediate step we shall first find the correlator on the plane. Afterwards, we will then use standard conformal transformation to get the desired cylinder result. We therefore start with a correlator of the form
\begin{equation}\label{eq:n11n}
\langle\sigma_{(1\cdots n)}(\infty)\bfO^{(j_1)}(1)\bfO^{(j_2)}(v)\sigma_{(n\cdots 1)}(0)\rangle\,,\qquad j_{1,2}\in\{1,\cdots n\}\,,
\end{equation}
which can be computed by going to the $n$-fold cover space using the map
\begin{equation}
z=t^n.
\end{equation}
The cover map undoes the twists inserted at $z=0$ and $z=\infty$\, and maps the two operator insertions at $z=(1,v)$ to $t=(t_1,t_2)=\left(e^{2i\pi j_1/n},v^{1/n}e^{2i\pi j_2/n}\right)$. On the cover space \eqref{eq:n11n} becomes a simple two-point function of seed primary operators:
\begin{equation}\label{eq:cover2pt}
\left\langle \bfO_{\text{seed}}(t_1)\bfO_{\text{seed}}(t_2)\right\rangle=\frac{1}{(t_1-t_2)^{2h_\bfO}}\,.
\end{equation}
Here, and in the following steps, we will focus on just the left-moving part of the correlator; the right-moving contribution can simply be multiplied at every step of the computation.
We can invert the map to get back to the correlator on the base space:
\begin{equation}
   \! \langle\sigma_{(1\cdots n)}(\infty)\bfO^{(j_1)}(1)\bfO^{(j_2)}(v)\sigma_{(n\cdots 1)}(0)\rangle = 
    \frac{1}{n^{2h_\bfO}v^{h_\bfO}\left(e^\frac{i\pi (j_1-j_2)}{n}v^{\frac{1}{2n}}-e^\frac{i\pi (j_2-j_1)}{n}v^{\frac{-1}{2n}}\right)^{2h_\bfO}}\,,
\end{equation}
with the $n^{2h_\bfO}$ coming from the Jacobian needed for the conformal transformation of the two probe operators. 

After going to the cylinder via $z=e^{-iw}$ we obtain
\begin{equation}\label{eq:twistncor}
\langle\sigma_{(1\cdots n)}(\infty)\bfO^{(j_1)}(0)\bfO^{(j_2)}(w)\sigma_{(n\cdots 1)}(-\infty)\rangle=\frac{1}{\left(2n\sin\left(\frac{w+2\pi(j_1-j_2)}{2n}\right)\right)^{2h_\bfO}}\,.
\end{equation}
Finally, by reinstating the symmetrization in \eqref{eq:untwOO} and including the anti-holomorphic part of the correlator, we arrive at 
\begin{equation}\label{eq:untwResult}
    \langle\sigma_{n}(\infty)\bfO(0)\bfO(w,\bar{w})\sigma_{n}(-\infty)\rangle_{\rm c}=\frac{n}{N}\sum_{j=0}^{n-1}\frac{1}{\left(2n\sin\left(\frac{w-2\pi j}{2n}\right)\right)^{2h_\bfO}\left(2n\sin\left(\frac{\bar{w}-2\pi j}{2n}\right)\right)^{2\bar{h}_\bfO}}\,.
\end{equation}
Here we have simplified the sum over $j_1$ and $j_2$ by realizing that only their difference $j\coloneqq j_1-j_2$ appears, leading to a factor of $n$.
As expected, this four-point function depends only on the dimension of the four operators and is independent of any seed OPE coefficients. Note that if the twist cycle is short, namely $n \sim N^0$ in the large $N$ limit, this result has the expected $N$ scaling consistent with large-$N$ factorization, see \eqref{eq:factorization}. Nevertheless, \eqref{eq:untwResult} is more general holds for any $2\leq n \leq N$. 

If we now take $n,N\rightarrow\infty$, with $\frac{n}{N}\leq1$ fixed, the twisted ground states become heavy and we get a heavy-heavy-light-light type correlator. In this limit \eqref{eq:untwResult} simplifies to 
\begin{equation}\label{eq:end4.1}
    \langle\sigma_{n}(\infty)\bfO(0)\bfO(w)\sigma_{n}(-\infty)\rangle_{\rm c}=\frac{n}{N}\sum_{j\in\mathbb{Z}} \frac{1}{\left(w-2\pi j\right)^{2h_\bfO}\left(\bar{w}-2\pi j\right)^{2\bar{h}_\bfO}}+\cdots\,.
\end{equation}
We see that in this limit there is a contribution to the connected four-point function \eqref{eq:4ptgen} that scales as $N^0$. At large $N$ the four-point function
\[
\langle\sigma_{n}(\infty)\bfO(0)\bfO(w)\sigma_{n}(-\infty)\rangle
\]
not only has a disconnected piece but also a connected contribution at leading order in $N$. We should stress though that this is not violating large-$N$ factorization, because the twist-2 insertions are heavy. The subleading contributions can be found by keeping additional terms in the Taylor expansion of the sine in \eqref{eq:untwResult}.

\subsubsection*{\refstepcounter{subsubsection}\label{sec:typ}\thesubsubsection\quad The ensemble of ground states}

Up until this point our explicit computations have been done for a very specific choice of twist-$[g]$ ground states $\sigma_{[g]}$, that is we have explicitly computed \eqref{eq:singlecycle4pt}. Here we discuss how this explicit computation also contributes to more general choices for $[g]$. Recall that a twisted ground state is fully characterized by an integer partition of integers $\{N_n\}_{n\in\mathbb{N}}$
\eq{\label{eq:part}
N=\sum_{n=1}^\infty n N_n\,.
}
As $N$ grows large, the number of partitions grows as
\eq{
\#\text{ partitions of }N\sim e^{2\pi\sqrt{N}}\,.
} 
We can then ask about the behavior of \eqref{eq:4ptgen} for different choices of those partitions. Do most partitions lead to a similar result? Or in other words, how universal are the answers we found in the previous section? To analyze such questions we need to understand better what a putative typical partition of $N$ would look like. A useful way to think about this problem is by noticing that we can think about a partition as a system of indistinguishable particles with energy levels $n\in\mathbb{Z}_{\ge0}$. A partition of $N$ then corresponds to a configuration of such particles with occupation numbers $N_n$ for the energy levels $n$ with total energy $N$ in accordance with \eqref{eq:part}. Ideally, we would fix the total energy $N$ to some (very large) value, that is, to study the microcanonical ensemble of the system at some large energy $N$. This is however very cumbersome. It is easier to work in the canonical ensemble instead, and the answers will agree since the ensembles are equivalent in the thermodynamic limit, i.e. at large $N$. We thus introduce an inverse temperature $\beta$, which should not be confused with any physical temperature; it is merely the conjugate variable of $N$. The canonical partition function at inverse temperature $\beta$ of this system is closely related to the Dedekind-eta function
\eq{
Z(\beta)=\Tr\left[e^{-\beta N}\right]=\prod_{n=1}^\infty\left(1-q^n\right)^{-1}=q^{\frac{1}{24}}\eta(\tau)^{-1}\,,\qquad\text{where}\qquad q\coloneqq e^{2\pi i\tau}=e^{-\beta}\,.
}
The relation between $N$ and $\beta$, at small $\beta$, can be found by using the modular properties of the Dedekind-eta function
\eq{
Z(\beta)\sim\sqrt{\frac{\beta}{2\pi}}e^{\frac{\pi^2}{6\beta}}\,,
}
so that
\eq{
\left\langle N\right\rangle=-\partial_\beta\log Z(\beta)\approx \frac{\pi^2}{6\beta^2}\,.
}
We thus find that for small $\beta$ we have $\beta\sim\frac{1}{\sqrt{N}}$, and hence $\beta\ll1$ is the regime relevant for large $N$. Via the analogy with indistinguishable particles, it follows that the occupation numbers $N_n$ follow the Bose-Einstein distribution
\eq{\label{eq:BE}
\langle N_n\rangle = \frac{1}{e^{\beta n}-1}\,.
}
We can infer a lot from the expectation value for the occupation numbers. For one, we can understand the variance of the occupation numbers via
\eq{
\text{Var}(N_n)=\langle \langle N_n\rangle-N_n\rangle^2=-\frac{1}{n}\frac{\partial\langle N_n\rangle}{\partial\beta}=\langle N_n\rangle(1+\langle N_n\rangle)\,.
}
 In particular, this implies that the difference between individual partitions is not necessarily small, since
\eq{
\frac{\sqrt{\text{Var}(N_n)}}{\langle N_n\rangle}=\mathcal{O}\left(\langle N_n\rangle^0\right)\,.
}

Here, we would like to clarify and correct some statements found in the literature. A variance being $\mathcal{O}(1)$ means, by definition, that there is no typical partition. Strictly speaking, some statements made in \cite{Balasubramanian:2005qu,Balasubramanian:2008da,Balasubramanian:2016ids} are thus incorrect. However, the final results in those papers are still correct. This is because even though there is no typical partition, the normalized variance does not scale with $N$. This means that while it does not make sense to ask if a generic partition has 2 or 6 cycles of size $\sqrt{N}$, it does make sense to say that a generic partition cannot have $\sqrt{N}$ cycles of size $\sqrt{N}$. In other words, while there is no typical partition, there \textit{is} a typical $N$ scaling of the partitions. In the end, this is really what is important in \cite{Balasubramanian:2005qu,Balasubramanian:2008da,Balasubramanian:2016ids} and explains why their results are correct even if there is no typical partition.

For a general partition, we will consider untwisted probes, i.e., we are in the setup of Sec.\,\ref{sec:notwist}. Since at most one nontrivial cycle can contribute to a connected term in the correlation function we have
\eqsp{
\langle\sigma_{[g]}(\infty)\bfO(0)\bfO(w)&\sigma_{[g]}(-\infty)\rangle \\
&=\frac{1}{N}\sum_{n=1}^\infty n N_n\sum_{j=1}^n\langle\sigma_{(1\cdots n)}(\infty)\bfO^{(1)}(0)\bfO^{(j)}(w)\sigma_{(n\cdots 1)}(-\infty)\rangle_{\rm c}\\
&=\frac{1}{N}\sum_{n=1}^\infty n N_n\sum_{j=1}^n\frac{1}{\left(2n\sin\left(\frac{w-2\pi j}{2n}\right)\right)^{2h_\bfO}\left(2n\sin\left(\frac{\bar{w}-2\pi j}{2n}\right)\right)^{2\bar{h}_\bfO}}\,.
}
This result is valid and exact for any choice of partition $\{N_n\}$. We will now study what happens in the ensemble over all partitions. We will now replace the occupation numbers $N_n$ with their expectation value. Again, we stress that this can have an order one variance, but if we are only looking for $N$-dependent scalings, then the variance does not affect the answer. For every $n\lesssim\sqrt{N}$ we have
\eq{
n\langle N_n\rangle\sim\sqrt{N}\,,
}
at large $N$, whereas for larger values of $n$ the occupation number is exponentially suppressed in some positive power of $N$. Those terms therefore do not contribute to the sum over $n$. In principle, every $n\lesssim\sqrt{N}$ thus contributes equally to the correlation function, and we obtain 
\eq{
\frac{1}{\sqrt{N}}\sum_{n=1}^{\sqrt{N}}\sum_{j=1}^n\frac{1}{\left(2n\sin\left(\frac{w-2\pi j}{2n}\right)\right)^{2h_\bfO}\left(2n\sin\left(\frac{\bar{w}-2\pi j}{2n}\right)\right)^{2\bar{h}_\bfO}}\,.
}
We now note that for any $\gamma<\frac{1}{2}$ the number of terms satisfying $n\le N^\gamma$ goes to zero relative to $\sqrt{N}$. We can thus approximate the sum by the terms where $n$ scales with $N^{1/2}$.\footnote{For example, we can split the sum between terms with $n$ bigger or smaller than $N^{1/4}$. The terms below give a vanishing contribution in the large $N$ limit, while the ones above combine to give \eqref{eq:typicaluntw}.} There are $\mathcal{O}(\sqrt{N})$ such terms, and in each of them, we can approximate the sine by a linear function. In the end, we arrive at 
\eq{\label{eq:typicaluntw}
\langle\sigma_{[g]}(\infty)\bfO(0)\bfO(w)\sigma_{[g]}(-\infty)\rangle=\sum_{j\in\mathbb{Z}} \frac{1}{\left(w-2\pi j\right)^{2h_\bfO}\left(\bar{w}-2\pi j\right)^{2\bar{h}_\bfO}}+\cdots\,,
}
with the $\cdots$ denoting contributions suppressed in $N$. For $g$ characterized by a partition with typical $N$ scaling, we thus find a universal result, that is also identical to \eqref{eq:untwResult} (with $n\rightarrow N$). Hence, we have found a large amount of universality. In the following, we investigate whether this universality extends to twisted sector probe operators.

\subsection{Twisted probes}\label{sec:addtwist}

We would now like to repeat the analysis with a twisted probe operator. For $\bfOp$, we will consider a twist-two excited state, generated by the primary seed operator $\bfOs$, which we denote by $\bfO_{\sigma_2}$. This operator has scaling dimension $h_{\bfO,2}$, with
\begin{equation}\label{eq:hndef}
    h_{\bfO,n} = \frac{h_{{\bfO}}}{n} + \frac{c}{24}\left(n-\frac{1}{n}\right).    
\end{equation}
The first piece comes from the primary excitation, with $h_\bfO$ being the dimension of $\bfOs$, and the second from the twist-$n$ field (which for $\bfOp$ we will take to be $n=2$).

The types of cycles in the twist-$[g]$ operators that can contribute to the connected component of \eqref{eq:4ptgen} are now more intricate. Because the probe operators are inserted on two copies (each) at the same time, at most four cycles of the twist-$g$ operators can be involved. 

For simplicity we will restrict ourselves to the case where both twist-$[g]$ operators have two cycles of identical length that contribute. That is, we will consider \eqref{eq:4ptgen} with the choice $g=(1\cdots n)(n+1\cdots 2n)$, for some $n>2$. With a slight abuse of notation, we will write $\sigma_{[g]}=\sigma_{(n,n)}$. We will thus evaluate
\begin{equation}
\langle\sigma_{(n,n)}(\infty)\bfO_{\sigma_2}(0)\bfO_{\sigma_2}(w)\sigma_{(n,n)}(-\infty)\rangle_{\rm c}\,.
\end{equation}
Depending on which copies the twist-2 insertions lie, the relevant cover surface can be of genus zero or of genus one, so that
\eqsp{\label{eq:genus}
\langle\sigma_{(n,n)}(\infty)\bfO_{\sigma_2}(0)\bfO_{\sigma_2}(w)\sigma_{(n,n)}(-\infty)\rangle_{\rm{c}}=&\langle\sigma_{(n,n)}(\infty)\bfO_{\sigma_2}(0)\bfO_{\sigma_2}(w)\sigma_{(n,n)}(-\infty)\rangle_{\rm{c},0}\\
&+\langle\sigma_{(n,n)}(\infty)\bfO_{\sigma_2}(0)\bfO_{\sigma_2}(w)\sigma_{(n,n)}(-\infty)\rangle_{\rm{c},1}\,,
}
where we have split the correlator according to the genus of the cover surface, and denoted it with a subscript. The genus zero cover arises when the twist-2 insertion acts on the two cycles of length $n$ by joining them to a cycle of length $2n$. The genus one cover on the other hand comes about when the twist-2 insertion acts by breaking up one of the length $n$ cycles into two smaller ones. We will focus on the genus zero contribution, and show how to compute it explicitly since this piece is universal. The genus one contribution is not universal. We will comment at the end of this section on when the genus zero terms dominates over the genus one contribution at large $N$.

In this setup, the combinatorics involved in writing \eqref{eq:4ptgen} as a sum over correlation functions of operators in fixed copies are more involved. Nevertheless, they can be worked out and lead to
\begin{align}\label{eq:fixedtwist}
&\langle\sigma_{(n,n)}(\infty)\bfO_{\sigma_2}(0)\bfO_{\sigma_2}(w)\sigma_{(n,n)}(-\infty)\rangle_{\rm{c},0}\\
&\hspace{20pt}=\frac{2n}{N(N-1)}\sum_{a=n+1}^{2n}\langle\sigma_{(1\cdots n)(n+1\cdots2n)}(\infty)\bfO_{\sigma_{(1a)}}(0)\bfO_{\sigma_{(1a)}}(w)\sigma_{(n\cdots 1)(2n\cdots n+1)}(-\infty)\rangle\,.\notag
\end{align}
As was the case in Sec.\,\ref{sec:notwist}, it is convenient to first consider the correlation functions on the plane with coordinate $v$, which is related to $w$ through $v=e^{-iw}$. We thus consider
\eq{
\langle\sigma_{(n,n)}(\infty)\bfO_{\sigma_2}(1)\bfO_{\sigma_2}(v)\sigma_{(n,n)}(0)\rangle_{\rm{c},0}\,.
}

As before, we map the twisted correlator to a cover space that undoes the twists, now located at $0,v,1,$ and $\infty$. For variables $z$ on the base and $t$ on the cover such a map is given by \cite{Lima:2020nnx}
\begin{equation}\label{eq:nncovermap}
    z(t) = \left(\frac{t}{t_1}\,\frac{(t-t_0)}{(t_1-t_0)}\,\frac{(t_1-t_\infty)}{(t-t_\infty)}\right)^n\,,
\end{equation}
where
\begin{equation}\label{eq:nnrel}
    t_0 =x-1\,, \quad t_1 = \frac{x(x-1)}{x+1}\,,\quad \text{and}\quad t_\infty = \frac{x^2}{x+1}\,,
\end{equation}
such that
\begin{equation}\label{eq:nncovering}
    v \coloneqq z(x) = \left(\frac{x+1}{x-1}\right)^{2n}\,.
\end{equation}
Under this map, the two cycles of length $n$ at $0$ and $\infty$ are separated and mapped to $t=0,t_0$ and $t=t_\infty,\infty$, respectively. Note that this cover map introduces factors of two that will later need to be accounted for since the symmetry between the two cycles of length $n$ is broken by mapping them to separate points on the cover space. The twist-two insertions of the probe operator at $1$ and $v$ are mapped to $t=1$ and $t=x$.

As a check of the validity of the cover map \eqref{eq:nncovermap} we can expand it around the different insertions and recover the expected behavior \eqref{eq:covermap}:
\begin{equation}
        z-z_* = b_{t_*}(t-t_*)^{n_*} + \cdots\,.
\end{equation}
In particular, we have
\begin{align}\label{eq:bxt1}
    \begin{split}
        &z_* = 1\,, t_*=t_1: \quad b_{t_1}= -\frac{n}{x}\left(\frac{x+1}{x-1}\right)^2\,,\\
        &z_* = v\,, t_*=x: \quad b_{x}= \frac{n}{x}\left(\frac{x+1}{x-1}\right)^{2n}\,,\\
    \end{split}
\end{align}
for the twist-two insertions with $n_*=2$.

The map \eqref{eq:nncovering} has inverses of the form
\begin{equation}\label{eq:nnsolutions}
    x_\alpha(v)= -\frac{1+v^{\frac{1}{2n}}e^{\frac{\alpha\pi i}{n}}}{1-v^{\frac{1}{2n}}e^{\frac{\alpha\pi i}{n}}}\,, \qquad \text{for}\qquad \alpha \in\{0,1,2,\dots,2n-1\}\,.
\end{equation}
There are $2n$ solutions, labeled here by $\alpha$, that can be grouped into $n$ pairs by noticing that
\begin{equation}\label{eq:xarelation}
    x_{\alpha+n}=\frac{1}{x_\alpha}\,,\qquad\text{for}\qquad \alpha\in\{0,\dots, n-1\}\,.
\end{equation}
Each pair of solutions corresponds to one value of $a$ in \eqref{eq:fixedtwist}. That is, we can identify $a$ with the solutions $x_{a-1}$ and $x_{a-n-1}$. Hence, by summing over the inverses of the cover map we can implement the sum over $a$ in \eqref{eq:fixedtwist}.

On the cover space, the remaining correlator reduces to a two-point function of $\bfOs$, precisely as was the case for the untwisted probe. We will denote it as
\begin{equation}
    G(x)|_{\text{cover}} = \left\langle \bfO_{\text{seed}}(t_1)\bfO_{\text{seed}}(x)\right\rangle = \frac{1}{(t_1-x)^{2h_\bfO}}\,.
\end{equation}
To determine the base correlator $G_\alpha(v)$, where now we have added a label $\alpha$ denoting which inverse \eqref{eq:nnsolutions} we use to map back to the base, we proceed following the method of Lunin and Mathur \cite{Lunin:2000yv,Lunin:2001pw}, see also \cite{Lima:2021wrz}. The correlator on the base in terms of the cross-ratio on the cover, $x$,  is given by 
\begin{equation}\label{eq:GxLM}
    G_\alpha(x)|_{\text{base}} = \mathcal{N} e^{S_L} b_{t_1}^{-\frac{h_\bfO}{2}}b_{x}^{-\frac{h_\bfO}{2}} G_\alpha(x)|_{\text{cover}},
\end{equation}
where the factors of $b_{t_*}$ come from the (regularized) transformation of the primary and are given in \eqref{eq:bxt1}, and $S_L$ denotes the Liouville action capturing the bare twist contribution. Moreover, we must include a normalization factor $\mathcal{N}$ due to regularization factors in the Liouville action. The normalization factor $\mathcal{N}$ can be fixed by taking OPE limits once we know the functional dependence of \eqref{eq:4ptgen}. For the cover map \eqref{eq:nncovermap}, the Liouville action $S_L$ is given by
\begin{equation}
    S_L = \frac{c}{8}\left((1+n)\log\!\left(x-1\right)+(1-n)\log\!\left(1+x\right)-\log\!\left(x\right)\right).
\end{equation}
We now have all the ingredients to evaluate the correlator \eqref{eq:GxLM} on the base space. Using the relation of $t_1$ with $x$ from \eqref{eq:nnrel}, we find
\begin{align}\label{eq:gvac}
\begin{split}
    G_\alpha(x)|_{\text{base}} &= \mathcal{N} \frac{(1-x)^{(1+n)\left(\frac{c}{8}+h_\bfO\right)}(x+1)^{(1-n)\left(\frac{c}{8}+h_\bfO\right)}}{(4n)^{\frac{h_\bfO}{2}}x^{\left(\frac{c}{8}+h_\bfO\right)}}\\
    &=\mathcal{N} \frac{1}{(4n)^{\frac{h_\bfO}{2}}v^{h_{\bfO,2}}}\left(\frac{1}{x}-x\right)^{2h_{\bfO,2}},
    \end{split}
\end{align}
where in the second line we used \eqref{eq:hndef} and \eqref{eq:nncovering} to simplify the expression. 

At this point in the computation, we cannot fix the normalization factor $\mathcal{N}$ completely, since it is fixed by taking OPE limits of the {\it full} correlator \eqref{eq:4ptgen}. We can however split the normalization factor in two pieces. One piece that normalizes $G_\alpha$, up to a factor $\mathcal{N}_0$, and a piece that we call $\mathcal{N}_0$ which can only be determined by taking OPE limits of the correlation function \eqref{eq:4ptgen}. To this end, we take the limit $v\rightarrow 1$ in \eqref{eq:gvac}, that is, we take the limit where the two probe operators approach each other. The cover map \eqref{eq:nncovering} has two inverses in this limit:
\begin{align}
    \begin{split}
        x\rightarrow \infty\,, \quad x(v)&=-\frac{4n}{1-v}+\cdots\,,\\
        x\rightarrow 0\,, \quad x(v)&=\frac{1-v}{4n}+\cdots\,.\\
    \end{split}
\end{align}
Expanding \eqref{eq:gvac} around either limit gives
\begin{equation} G_\alpha(x\rightarrow\infty\,\text{or} \,0)|_{\text{base}}=\frac{\mathcal{N}\left(4n\right)^{2h_{\bfO,2}-h_\bfO/2}}{(v-1)^{2h_{\bfO,2}}}+\cdots.
\end{equation}
This first term encodes the exchange of the identity operator and comes with unit normalization, hence we can deduce that $\mathcal{N}=\mathcal{N}_0\left(4n\right)^{\frac{h_\bfO}{2}-2h_{\bfO,2}}$.

As a check, we also consider the other OPE channel, $v\rightarrow 0$, in which the twist-2 field approaches the two twist-$n$ fields. In this limit, there is a single inverse of \eqref{eq:nncovering} that satisfies
\begin{equation}
    x\rightarrow -1\,, \quad x(v)=-1+2 v^{\frac{1}{2n}}+\cdots\,,
\end{equation}
and we find
\begin{equation}\label{eq:GxOPEv0}
G_{\alpha}(x\rightarrow -1)|_{\text{base}}=\mathcal{N}_0\frac{v^{h_{\bfO,2}\left(\frac{1}{n}-1\right)}}{n^{h_{\bfO,2}}}+\cdots.
\end{equation}
The first term corresponds to the exchange of a twist-$2n$ state with an $\bfOs$ excitation on top, as expected, since 
\begin{equation}
     h_{\bfO,2n} - 2h_{\mathbb{1},n} - h_{\bfO,2} = \left(\frac{c}{16}+\frac{h_\bfO}{2}\right)\left(\frac{1}{n}-1\right) = h_{\bfO,2}\left(\frac{1}{n}-1\right)\,.
\end{equation}

Since we are interested in correlators on the cylinder, we insert the inverse of the cover map \eqref{eq:nnsolutions} into \eqref{eq:gvac} to get back to the base space variable $v$ and then map to the cylinder via $v=e^{-iw}$:
\begin{equation}\label{eq:Gabare}
    G_\alpha(w) = \frac{\mathcal{N}_0}{\left(2n \sin\!\left(\frac{2\alpha\pi-w}{2n}\right)\right)^{2h_{\bfO,2}}}\,.
\end{equation}
The final step we need to take in order to evaluate \eqref{eq:fixedtwist} is to relate $G_\alpha(w)$ to the terms in the sum. As we already mentioned, the $G_\alpha$'s have a symmetry that is related to the exchange of the two twist cycles of length $n$ in the ground state insertions. This symmetry gives rise to a factor two, and the term in \eqref{eq:fixedtwist} labeled by $a$ is equal to
\eq{
\frac{1}{2}\left(G_a(w)\bar{G}_a(\bar{w})+G_{a-n}(w)\bar{G}_{a-n}(\bar{w})\right)\,,
}
where with a bar we denote the anti-holomorphic version of $G_a$. Hence, the sum over the different inverses of the cover map is equivalent to the sum over $a$ in \eqref{eq:fixedtwist}! Combining the expressions we thus obtain
\begin{align}\label{eq:twResult}
    \begin{split}
    &\hspace{-40pt} \langle\sigma_{(n,n)}(\infty)\bfO_{\sigma_2}(0)\bfO_{\sigma_2}(w)\sigma_{(n,n)}(-\infty)\rangle_{\rm{c},0}\\
     &=\frac{n}{N(N-1)}\sum^{n-1}_{\alpha=0}\left(G_\alpha(w)\bar{G}_\alpha(\bar{w})+G_{\alpha+n}(w)\bar{G}_{\alpha+n}(\bar{w})\right)\\
        &=\mathcal{N}_0^2\frac{2n}{N(N-1)}\sum^{n-1}_{j=0}\frac{1}{\left(2n\sin\!\left(\frac{2j\pi-w}{2n}\right)\right)^{2h_{\bfO,2}}\left(2n\sin\!\left(\frac{2j\pi-\bar{w}}{2n}\right)\right)^{2\bar{h}_{\bfO,2}}}\,,
    \end{split}
\end{align}
where we have used $G_{\alpha+n} = (-1)^{2h_{\bfO,2}}G_\alpha$ by inserting \eqref{eq:xarelation} in \eqref{eq:Gabare}. Furthermore, we note that, up to a combinatorial factor, this expression agrees with the result for untwisted probes \eqref{eq:untwResult}. We also remark that the $N$ scaling of the combinatorial factor is consistent with large-$N$ factorization.

In the large-$N$, large-$n$ limit, which we take by keeping $\frac{n}{N}\le\frac{1}{2}$ fixed, the twisted ground state operators become heavy, and \eqref{eq:twResult} simplifies to
\eqsp{\label{eq:end4.2}
&\langle\sigma_{(n,n)}(\infty)\bfO_{\sigma_2}(0)\bfO_{\sigma_2}(w)\sigma_{(n,n)}(-\infty)\rangle_{\rm{c},0}\\
&\hspace{100pt}=\mathcal{N}_0^2\frac{2n}{N(N-1)}\sum_{j\in\mathbb{Z}} \frac{1}{\left(w-2\pi j\right)^{2h_\bfO}\left(\bar{w}-2\pi j\right)^{2\bar{h}_\bfO}}+\cdots\,.
}
This expression comes about in exactly the same way as \eqref{eq:end4.1}, namely by Taylor expanding the sines in \eqref{eq:twResult}. Moreover, the functional dependence is identical to the functional dependence of \eqref{eq:end4.1}. However, we cannot determine the scaling with $N$ since we did not fix the normalization factor $\mathcal{N}_0$.  $\mathcal{N}_0$ depends on $n$, and is fixed by taking into account the contributions from the genus one term in \eqref{eq:genus}. 


\subsubsection*{\refstepcounter{subsubsection}\thesubsubsection\quad The ensemble of ground states \label{sec:HHLL}}

Just as we discussed in Sec.\,\ref{sec:typ}, we can wonder whether and how our explicit results for twist-2 probes generalize to other, more general, choices of $g$. Recall that $[g]$ is fully characterized by the integer partition
\eq{\label{eq:part2}
N=\sum_{n=1}^\infty n N_n\,.
}
As we explained, for twist-2 probes there are two qualitatively different contributions to \eqref{eq:4ptgen}. First, the two copies involved in the probe insertions can overlap with two different cycles in $\sigma_g$. This leads to a cover surface of genus 0, as can be seen from the Riemann-Hurwitz formula. The second type of contribution appears when the copies of the twist-2 probe overlap with a single cycle of $\sigma_g$. In that case, the cover surface is a torus. How do these different contributions scale? It can be shown that the torus contribution is multiplied by a factor that scales as $N_n$, whereas the genus zero contribution comes with a scaling $N_{n_1}N_{n_2}$, with $n_1$ and $n_2$ the length of the cycles that the probe operator joins together.\footnote{When the cycle lengths of the cycles that are joined together are equal, i.e., when $n_1=n_2$ the correct factor to multiply with is $\binom{N_n}{2}$. The intuition is that out of the $N_n$ cycles of length $n$, we pick out two that are joined together by the twist-2 probe.} It follows that when the occupation numbers are large, the genus zero contribution dominates. In particular, this is true for partitions following the Bose-Einstein distribution. Therefore, our focus on the genus zero contribution in \eqref{eq:genus} is justified when the occupation numbers are large.

The specific choice of $g$ we made, $g=(1\cdots n)(n+1\cdots 2n)$, clearly has $N_n=2$. This means that the genus zero and genus one contribution in \eqref{eq:genus} are equally important, regardless of the length of the cycles. Hence, in order to fix the normalization in our final result \eqref{eq:end4.2}, we really need the genus one contribution. There exist explicit expressions for the cover map to the torus, but they are not invertible, see e.g. \cite{Lunin:2000yv}. That means that there are no known expressions for the analog of $x_\alpha$ in \eqref{eq:nnsolutions}. It follows that, at least for this choice of $g$, we cannot fully pin down the four-point function \eqref{eq:4ptgen}, even in the limit where $n$ and $N$ are large.

There is another choice for $g$ to which our computations apply straightforwardly, and for which we can completely fix the normalization, at least to leading order in $N$. Instead of considering two cycles of length $n$, we consider $q\coloneqq N/n$ cycles of length $n$, that is, we let
\eq{\label{eq:g2}
g=\underbrace{(1\cdots n)(n+1\cdots 2n)\cdots (N-n\cdots N)}_{q\text{ cycles}}\,.
}
With this choice of $g$ every possible contribution to \eqref{eq:4ptgen} will be proportional to one of the terms in \eqref{eq:genus}. In fact, we can work out the combinatorics to obtain
\eqsp{
\langle \sigma_{[g]}(\infty) \bfO (0) \bfO (w) \sigma_{[g]}(-\infty)\rangle&=\mathcal{N}_0^2\frac{2n}{N(N-1)}\Bigg[q\times\sum \text{genus 1 cover maps}\\
&\hspace{20pt}+ \binom{q}{2}\times\sum\text{genus 0 cover maps}\Bigg]\,.
}
Here the sums are over the different non-equivalent configurations of the non-gauge invariant insertions $\sigma_g$ and $\sigma_{(ab)}$, and we have taken out the normalization factor $\mathcal{N}_0$. In fact, the sum over genus zero cover maps is precisely the sum over $j$ in \eqref{eq:twResult}. We see that once we increase $q$ the genus zero contribution starts to dominate. As a particularly nice example, let us consider $n=q=\sqrt{N}$. The genus one contribution is suppressed by a factor $\sqrt{N}$ compared to the leading order behavior coming from cover surfaces of genus zero.  We can use \eqref{eq:end4.2} to find
\eq{
\langle \sigma_{[g]}(\infty) \bfO (0) \bfO (w) \sigma_{[g]}(-\infty)\rangle=\mathcal{N}_0^2\frac{\sqrt{N}-1}{N-1}\sum_{j\in\mathbb{Z}}\frac{1}{\left(w-2\pi j\right)^{2h_\bfO}\left(\bar{w}-2\pi j\right)^{2\bar{h}_\bfO}}+\cdots\,.
}
Requiring the usual coincidence limit then fixes the normalization to be $\mathcal{N}_0^2=\frac{N-1}{\sqrt{N}-1}$.\footnote{Note that $\mathcal{N}_0$ is only allowed to depend on properties the group elements involved: $g$ and $(12)$. The $N$ dependence of $\mathcal{N}_0$ is purely a result of taking $n$ to scale non-trivially with $N$.} Our final result (for this choice of $g$) is thus identical to \eqref{eq:typicaluntw}
\eq{
\langle\sigma_{[g]}(\infty)\bfO(0)\bfO(w)\sigma_{[g]}(-\infty)\rangle=\sum_{j\in\mathbb{Z}} \frac{1}{\left(w-2\pi j\right)^{2h_\bfO}\left(\bar{w}-2\pi j\right)^{2\bar{h}_\bfO}}+\cdots\,.
}
Note that this result generalizes to instances where $q\sim N^\alpha$, and thus $n\sim N^{1-\alpha}$ for finite values of $\alpha\in(0,1)$. In that case, the genus one terms are suppressed by a factor $N^{\alpha}$. As in Sec.\,\ref{sec:notwist} we find a universal answer. 

We can even consider more general choices of $g$, namely permutations with nontrivial cycles of different lengths. For such choices of $g$ there is an additional genus zero contribution that one needs to take into account. Namely, the contribution where the twist-2 probe joins together two cycles of unequal length. In this case, to the best of our knowledge, there is no cover map that is invertible. Therefore, we are unable to explicitly compute contributions to \eqref{eq:4ptgen} of this form and hence, we cannot make statements about universality. 

It would be very interesting to understand generalizations to probe operators of higher twists. Naively, for a twist-$m$ probe, and a typical distribution for $[g]$, the dominating contribution comes from joining $m$ different cycles of $\sigma_{[g]}$ together, that is, the terms where the copies involved in the twist-$m$ overlap with $m$ distinct cycles of $\sigma_g$. The cover surface involved is again a sphere.
\\

We conclude this section by emphasizing that the correlator \eqref{eq:4ptgen} depends sensitively on the partition of integers associated with the conjugacy class $[g]$ and the probe operator $\bfOp$. Nevertheless, our results never depend on seed OPE coefficients, and for certain classes of choices, the functional dependence of \eqref{eq:4ptgen} is identical. This is illustrated by the specific examples considered in Sec.\,\ref{sec:notwist} and Sec.\,\ref{sec:addtwist}, and remains true as long as the partition of $N$ chosen is such that the contributions computed in Sec.\,\ref{sec:notwist} and Sec.\,\ref{sec:addtwist} dominate in the large-$N$ limit.

\section{Discussion}\label{sec:grav}

We conclude by summarizing our results regarding universality in correlation functions of symmetric product orbifolds at large $N$ in Sec.\,\ref{sec:5.1}. In the process, we also discuss their gravitational interpretation. In Sec.\,\ref{sec:5.2} we comment on the implication of our results with respect to the emergence of spacetime and horizons, and discuss some interesting open questions.

\subsection{Gravitational interpretation}\label{sec:5.1}

In this work, we have studied the universality of correlation functions in the large-$N$ limit of symmetric product orbifolds. A striking result of our analysis is complete universality of thermal two-point functions in the large-$N$ limit. Above the Hawking-Page transition, the thermal two-point function of untwisted single-trace scalar operators takes the form
\eq{\label{eq:sec5th}
\left\langle\bfO(t_{\rm E},\phi)\bfO(0)\right\rangle_\beta=
    \displaystyle\sum_{j\in\mathbb{Z}}\frac{1}{\left(\frac{\beta^2}{4\pi^2}\left(\cosh\left(\frac{2\pi(\phi+2\pi j)}{\beta}\right)-\cos\left(\frac{2\pi t_{\rm E}}{\beta}\right)\right)\right)^{2h_\bfO}}\qquad\text{for}\qquad\beta<2\pi\,.
}
This expression can be recognized as the boundary two-point function of scalar fields in a BTZ geometry with inverse temperature $\beta$ \cite{Keski-Vakkuri:1998gmz,Maldacena:2001kr,Balasubramanian:2005qu}. Here $h_\bfO$ is the scaling dimension of the CFT operator dual to the bulk scalar field. A similar result holds for the thermal AdS phase when $\beta>2\pi$. We emphasize that this result is valid for any choice of seed theory, regardless of whether the resulting symmetric product orbifold has a moduli space with a holographic point. Even for symmetric product orbifolds that do not have a coupling that, when turned on, leads to a strongly coupled theory, the thermal two-point function is equal to that of BTZ.

We also obtain universal results for the thermal one-point function which are consistent with the results above, large-$N$ factorization, and expectations from gravity. There are no primary fields whose one-point functions grow with $N$. On the gravity side, this implies that light fields do not cause a large backreaction on the geometry at finite temperatures, which would cause a deviation from the BTZ geometry. Our result for symmetric product orbifolds should be contrasted with other free or weakly-coupled large-$N$ gauge theories (like $\mathcal{N}=4$ SYM) where it is expected that all the one-point functions of single-trace operators are large \cite{Grinberg:2020fdj}. It is interesting to note that the result for our one-point function takes the form
\be\label{eq:5.2}
\braket{O}_\beta \sim \sqrt{N} e^{-\frac{N}{\beta}  \Delta_O} \,.
\ee
The first factor grows with $N$, and corresponds to the naive large $N$ counting expectation, namely that all one-point functions of single-trace operators should be order $\sqrt{N}$ at weak coupling. The second factor is responsible for turning off the one-point functions and is special to two-dimensional CFTs, where one-point functions at infinite temperature vanish due to conformal symmetry. Here, the effective temperature of the seed correlator becomes infinite as $N$ goes to infinity.

Another important result of our analysis is the two-point function of light operators in twisted ground states as computed in Sec.\,\ref{sec:twisteduniv}. Our results can be summarized by
\eq{\label{eq:sec5gs}
\bra{\sigma}\bfO(1)\bfO(w)\ket{\sigma}\sim\sum_{j=0}^{n-1}\frac{1}{\left(2n\sin\left(\frac{w-2\pi j}{2n}\right)\right)^{2h_\bfO}\left(2n\sin\left(\frac{\bar{w}-2\pi j}{2n}\right)\right)^{2\bar{h}_\bfO}}\,.
}
Here $\sigma$ is the ground state of the twist-$[g]$ sector for a permutation $g$ that has cycles of length $n$. We explicitly computed this correlator for $\bfO$ in the untwisted and twist-2 sectors. Surprisingly, the right-hand side of \eqref{eq:sec5gs} reproduces exactly the bulk two-point function of a light field in AdS$_3$ with a conical defect of order $n$. In particular, if $n\sim N^\alpha$ for some $0<\alpha\leq 1$, we recover the two-point function in the $M=0$ BTZ geometry as we take the limit $N\to\infty$ \cite{Balasubramanian:2005qu}. Note that by considering (light) two-point functions in twisted ground states one can never reach energies high enough to be dual to massive BTZ black holes. This can be seen from the fact that for choices of $[g]$ with long cycles, the weight of the twisted ground state reaches at most $\Delta= cN/12$, which is still below the threshold of BTZ black holes. 

The reason why calculations in symmetric product orbifolds agree with the $M=0$ BTZ result is somewhat mysterious to us. The quantity we have computed does not dominate the canonical ensemble (or even the microcanonical ensemble for that matter). We have simply studied particular microstates and found a result consistent with a two-point function computed in a bulk geometry of AdS with a conical defect. Moreover, as before, this result is completely independent of the choice of seed theory, and hence independent of the presence of a coupling that can be turned on to flow to a holographic theory.

We conclude by outlining some of the consequences of these results from a holographic perspective and listing some future directions.

\subsection{Mirage of the horizon}\label{sec:5.2}

The most prominent lesson to draw from our results is that finding thermal correlators consistent with a black hole geometry is not sufficient to ensure a holographic CFT, or to ensure that the bulk theory has a geometric description with a horizon. Thermal two-point functions in the large-$N$ limit play an important role in the algebraic approach to studying the emergence of spacetime. The spectral density of the thermal two-point function of an operator $\bfO$ is tied to the emergence of type III$_1$ algebras above the Hawking-Page transition for the operator algebra associated to $\bfO$ \cite{Leutheusser:2021frk,Leutheusser:2021qhd}. It was shown in \cite{Gesteau:2024dhj} that if and only if the spectral density has some region of continuous support there is an emergent type III$_1$ algebra. Moreover, if the spectral density has continuous support on the entire real line, the causal depth parameter is infinite \cite{Gesteau:2024rpt}, see also \cite{Lashkari:2024lkt}.

It should come as no surprise that the BTZ thermal two-point function satisfies all proposed conditions for the emergence of spacetimes with a horizon. From the perspective of \cite{Gesteau:2024rpt} our universal results thus point to an emergent local description on a  BTZ spacetime that is valid along the radial bulk direction all the way to the horizon. We take our results as strong evidence that having a continuous spectral density for thermal two-point functions cannot possibly be a sufficient condition for the emergence of a geometric spacetime. In fact, symmetric product orbifolds are, as reviewed in Sec.\,\ref{sec:symuniversality}, in some sense as far away as possible from holographic CFTs (within the class of CFTs with a consistent large-$N$ limit, large-$N$ factorization and the sparseness condition). The most pressing question is to understand exactly what gives in this picture. We discuss some possible resolutions below.

\subsubsection*{Sparseness and higher spin operators}

We believe that the most probable resolution comes from the huge number of light fields present in symmetric product orbifolds, the number of which grows exponentially with the scaling dimension. The BTZ two-point function clearly shows that for each single-trace operator treated \textit{individually}, there is a type III$_1$ Von Neumann algebra. However, the type of full Von Neumann algebra of the CFT in this limit is related to the spectrum of the modular Hamiltonian, which knows about all fields. Even though the single-trace operators are strictly non-interacting in this limit, one still needs to resum an infinite number of them.

A promising idea to explore is the importance of higher spin operators. These operators are very important from a bootstrap perspective, and the presence of the infinite tower of higher spin operators is the explanation for the lack of Lyapunov behavior in the OTOC. In the case of supergravity on AdS$_5\times S^5$, there is also an infinite number of single-trace operators, but only of bounded spin. The growth of all light operators is also slower, growing like
\be
\rho_{\text{AdS}_5\times S^5}(\Delta)\sim e^{\Delta^{\frac{9}{10}}} \,,
\ee
instead of exponentially. It would be very interesting to understand if the important distinction in symmetric product orbifolds is the higher spin operators, the exponential growth of all operators, or both.

An interesting context where a sharp distinction between Einstein gravity and string theory was found is in the entanglement entropy. The half-space entanglement entropy of bosonic string theory was found, within a particular scheme of analytic continuation, to be free of UV divergences in \cite{Dabholkar:2023ows}. Similar results were obtained in a context more relevant for us: the 1-loop string partition function about the BTZ background appears to be free of the same UV divergences \cite{Dabholkar:2023tzd}. These results, taken at face value, would imply that the resummation of all the stringy operators leads to a type I Von Neumann algebra, even if the Von Neumann algebra related to each individual single-trace operator is of type III. To the best of our knowledge, there is no EFT-like reduction of the string theory calculation to understand how an infinite number of infinite contributions gives rise to something finite. From the calculations of \cite{Dabholkar:2023ows}, the finiteness comes from cutting out the UV-sensitive regions in the modular integrals of string theory, but it is not clear at a practical level how this implements the resummation described above. It would be interesting to understand this better.

\subsubsection*{1/\textit{N} corrections.}

Another possible resolution regards the way the large-$N$ limit is taken. Both in this paper and in the algebraic requirements put forth, only the strict $N\rightarrow\infty$ data is discussed. Perhaps, in order to precisely quantify conditions for the emergence of a local bulk geometry with a horizon, one needs to take into account $1/N$ corrections, and the leading large-$N$ contribution to thermal two-point functions is not sufficient to distinguish a sharp horizon in the bulk. This resonates with explicit computations in fuzzball geometries, see e.g. \cite{Bena:2019azk}. From the algebraic perspective taking into account $1/N$ corrections can change the nature of the operator algebras. For example, entropies can be defined in a meaningful way \cite{Witten:2021unn}, and these modified algebras were recently used to obtain an operator algebra understanding of the black hole information problem \cite{vanderHeijden:2024tdk}.

It would be very interesting to understand the $1/N$ corrections to the universal correlation functions of symmetric product orbifolds computed in this work. Analogous to the partition function, the $1/N$ corrections are not expected to agree with those expected in a gravitational theory, as presumably, the details of the seed theory become important. One possible avenue to understanding subleading corrections is via crossing kernels, which have been used to derive asymptotic expansions to the dynamics of heavy operators and were recently used to quantify the subleading corrections to the elliptic genus in sparse supersymmetric CFTs \cite{Collier:2019weq,Belin:2021ryy,Anous:2021caj,deBoer:2024mqg,Apolo:2024kxn}. It is also worth mentioning that the study of chaos and OTOCs (which also probes the structure of horizons) is intrinsically a $1/N$ endeavor since it involves connected four-point functions that vanish in the strict large-$N$ limit.

\subsubsection*{Thermal correlators of twisted operators.}

The final possible resolution has to do with twisted operator thermal correlation functions. If different operators lead to different values of the associated causal depth parameter of \cite{Gesteau:2024rpt}, the interpretation should be that of a stringy geometry, where different probes see different geometries: a violation of the equivalence principle. Our results on thermal correlation functions in symmetric product orbifolds are valid for untwisted sector operators, for which we find universal results. The arguments we used to derive the results in the untwisted sector do not generalize to twisted sectors, because the cover surfaces generated when considering twisted insertions on the torus are of higher genus. It would be very interesting to investigate whether, despite these complications, the universality we found prevails even for twisted sector operators. There are two signs that indeed this might be the case. The first hint is the fact that in our computations of two-point functions in twisted ground states our universal result \eqref{eq:sec5gs} is valid for both untwisted and twist-2 operators; the details of the probe operator do not matter. Second, we can also derive universality by following the criteria of \cite{Kraus:2017kyl}, which are valid for any light operator, regardless of the details of the theory (besides the HKS sparseness condition, large-$N$ factorization and some mild additional condition). We go through the criteria in symmetric product orbifolds in App.\,\ref{app:hks2pt}, expanding on \cite{Alexunpublished}, and find good indication that the criteria are met, which would also imply universal twisted correlators. 

 \section*{Acknowledgments}
We thank Elliott Gesteau, Per Kraus, Nima Lashkari, Hong Liu, Kyriakos Papadodimas, Antony Speranza, and Erik Verlinde for helpful discussions. We also thank the organizers and participants of the Physics Sessions Initiative meetings in 2023 where the ideas of this project started; in particular, we thank Hong Liu, Thomas Hartman, and Henry Maxfield for discussions. AB thanks Tom Hartman and Edgar Shaghoulian for discussions and unpublished collaboration on topics close to this project. SB is supported in part by the National Science Foundation grant PHY-2209700 and the Mani L. Bhaumik Institute for Theoretical Physics. AC has been partially supported by STFC consolidated grants ST/T000694/1 and ST/X000664/1. WK has been partially supported by the National Science Foundation under Grant NSF PHY-2210533 and by the Simons Foundation under Grant 681267.

\appendix 

\section{Dominance of twisted sectors}\label{app:longcycles}

In this appendix, we review the argument for a universal regime of Hagedorn growth in symmetric product orbifolds. This was first shown in \cite{Keller:2011xi} by considering the grand canonical partition function \eqref{eq:ZHecke}. Because we rely heavily on the formalism introduced by Bantay in the main text, we will use here that formalism to analyze the partition function \eqref{eq:bantay}, as was done in \cite{Belin:2014fna}. 

Recall Bantay's formula for the torus partition function in a permutation orbifold
\eq{\label{eq:bantayapp}
Z_{G_N}(\tau)=\frac{1}{\abs{G_N}}\sum_{gh=hg}\prod_{\xi\in O(g,h)}Z_\mathcal{C}(\tau_\xi)\,.
}
First, we will analyze terms in the sum with $(g)=1$. In this case, the remaining sum is over all $h\in G_N$. We write $h$ in terms of its cycle distribution $h=\displaystyle\prod_n(n)^{j_n}$, with $\displaystyle\sum_n nj_n=N$. There are now $j_n$ orbits $\xi_n$ of length $n$, with $\kappa_\xi=0$, $\lambda_\xi=1$ and $\mu_\xi=n$. Hence, the term corresponding to the choice $g=1$ and $h=\displaystyle\prod_n(n)^{j_n}$ in the partition function is
\eq{
\frac{1}{\abs{G_N}}\prod_{n=1}^N Z_\mathcal{C}(n\tau)^{j_n}\,.
}
The sum over different choices of $h$ implements the group average over $G_N$, and we reproduce the partition function of the untwisted sectors \cite{Belin:2014fna}. We can thus interpret different choices of $h$ in \eqref{eq:bantayapp} as implementing invariance under group permutations.

Next, we consider nontrivial choices for $g$, while we set $h=1$. We start by considering permutations of the form $g=(1)^{N-L}(L)$. Associated to this choice of $g,h$ there are $N-L+1$ orbits, all with $\mu_\xi=1$ and $\kappa_\xi=0$. The orbit corresponding to the length $L$ cycle has $\lambda_\xi=L$, while the other $N-L$ orbits have $\lambda_\xi=1$. Hence this term in the sum \eqref{eq:bantayapp} is
\eq{
Z_\mathcal{C}(\tau)^{N-L}Z_\mathcal{C}(\tau/L)\,.
}
For large $L\gg1$, this implies that we can use Cardy's formula on the contribution from the length $L$ cycle. In terms of the density of states, we obtain a contribution of the form\footnote{For simplicity we consider the scalar contribution here, but the argument can be easily generalized, as was done in \cite{Keller:2011xi}.}
\eq{
\rho\left(\Delta\right)^{N-L}\rho\left(L\Delta\right)\sim e^{2\pi\sqrt{\frac{cL(\Delta-cL/12)}{3}}}\,.
}
This expression is maximized by cycles of length $L=\frac{6\Delta}{c}$. Hence, the dominant contribution is
\eq{
e^{2\pi\sqrt{\frac{cL(\Delta-cL/12)}{3}}}\underset{\Delta=\frac{cL}{6}}{=} e^{2\pi \frac{cL}{6}}\,.
}
It can be shown that the other contributions do not combine to give rise to a faster growth. We thus find that a twist $L\gg1$ cycle leads to Hagedorn growth 
\be
e^{2\pi \Delta} \,,
\ee
at $\Delta=\frac{cL}{6}$. If the permutation group $G_N$ contains permutations of the form $(1)^{N-L}(L)$ for values of $L$ dense in $\{1,\cdots,N\}$ as $N$ grows large, the density of states is bounded from below by Hagedorn growth starting at energies of the order of the AdS length.

If $G_N$ contains permutations with arbitrary large cycles as $N\rightarrow\infty$, then there is a regime of Hagedorn growth sourced by the permutations with long cycles \cite{Belin:2014fna}. Consider such an element $g=(L_1)\cdots(L_n)$, where $n$ is finite as $N$ grows large. This means that (some of) the cycles grow with $N$. The contribution due to this $g$ equals
\eq{
\prod_{i=1}^n Z_\mathcal{C}\left(\frac{\tau}{L_i}\right)\,.
}
For the cycles that grow with $N$ we can again use the Cardy formula to estimate the density of states, and we obtain
\eq{
\prod_{i}\rho(L_i\Delta)\sim e^{2\pi \sqrt{c\left(\sum_i L_i\right)\Delta/3}}\underset{\Delta=\frac{cN}{12}}{\sim} e^{2\pi cN/6}\,,
}
and we find Hagedorn growth, but this time at $\Delta=cN/12$.

\section{Universal properties of thermal correlators based on sparseness criteria}\label{app:hks2pt}

In this appendix, we review the results of \cite{Kraus:2017kyl} and discuss how they apply to symmetric product orbifolds \cite{Alexunpublished}. The goal of \cite{Kraus:2017kyl} was to generalize the considerations of HKS to thermal one- and two-point functions. That is, to show that given some conditions on the light sector of the theory that we will review momentarily, the thermal one- and two-point functions have a universal phase structure consistent with HKS.

The setup is that of families of CFTs with a large central charge limit. The spectrum of such CFTs is divided into light, medium, and heavy states
\eq{
L= \{0\le\Delta\le\Delta_c\}\,,\qquad M=\{\Delta_c\le\Delta\le\frac{c}{12}\}\,,\qquad H=\{\Delta>\frac{c}{12}\}\,.
}
Here the cutoff $\Delta_c$ is fixed as $c\rightarrow\infty$. Once the large-$c$ expansion is taken, the cutoff $\Delta_c$ is sent to infinity. 

The starting point of the computations is the thermal two-point function obtained from cutting the path integral along two time slices that separate the operators $\bfO$, and inserting a complete set of states along the cuts
\eq{\label{eq:full2pt}
\left\langle \bfO(\phi,t)\bfO(0,0)\right\rangle_\beta=\frac{e^{\beta\frac{c}{12}}}{Z(\beta)}\sum_{A,B}e^{-\beta\Delta_A}e^{t\Delta_{AB}}e^{i\phi J_{AB}}\abs{\bra{A}\bfO\ket{B}}^2\,.
}
The strategy is then to split the different contributions to the sum according to $A,B\in L,M,H$. Under the conditions, which we discuss momentarily, the authors show that below the Hawking-Page transition, the sum is dominated by the contributions from the sums restricted to $A,B\in L$. Using modular covariance one then obtains a similar expression valid above the phase transition. The final result is
\begin{equation}\label{eq:kss2pt}
\begin{aligned}
\hspace{-17pt}    \left\langle \bfO(\phi,t)\bfO(0,0)\right\rangle_\beta\approx\begin{cases}\displaystyle\frac{e^{\beta\frac{c}{12}}}{Z(\beta)}\sum_{A,B\in L}e^{-\beta\Delta_A}e^{t\Delta_{AB}}e^{i\phi J_{AB}}\abs{\bra{A}\bfO\ket{B}}^2\,, & \beta>2\pi\\
\displaystyle\left(\frac{2\pi}{\beta}\right)^{2\Delta}\frac{e^{\beta'\frac{c}{12}}}{Z(\beta)}\sum_{A,B\in L}e^{-\beta'\Delta_A}e^{\frac{\beta'}{2\pi}\phi\Delta_{AB}}e^{\frac{2\pi}{\beta}it J_{AB}}\abs{\bra{A}\bfO\ket{B}}^2\,, & \beta<2\pi
\end{cases}
\end{aligned}
\end{equation}
with $\beta'=\frac{4\pi^2}{\beta}$, and crucially, the sums are constrained to just the light sector. This expression bears resemblance to a sum over bulk Witten diagrams on a thermal AdS geometry below the phase transition and a BTZ geometry above it. Using the result for the thermal two-point function the authors then find a similar expression for the thermal one-point function as a sum over light OPE coefficients
\eq{\label{eq:kss1pt}
\left\langle \bfO\right\rangle_\beta\approx\begin{cases}\displaystyle\frac{1}{Z(\beta)}\sum_{A\in L}e^{-\beta E_A}\bra{A}\bfO\ket{A}\,, & \beta>2\pi\\
\displaystyle\left(\frac{2\pi}{\beta}\right)^{2\Delta}\frac{1}{Z(\beta)}\sum_{A\in L}e^{-\frac{4\pi^2}{\beta}E_A}\bra{A}\bfO\ket{A}\,, & \beta<2\pi
\end{cases}\,~.
}

The results \eqref{eq:kss1pt} and \eqref{eq:kss2pt} are valid if the CFT satisfies four conditions, which we now discuss in turn.

\paragraph{Sparse spectrum.} The first condition is the HKS sparseness condition mentioned in Sec.\,\ref{sec:symuniversality}. Symmetric product orbifolds universally satisfy the HKS condition \cite{Keller:2011xi}.

\paragraph{Factorization of light correlators.} The second condition is that of large-$N$ factorization of light correlators in the large-$N$ limit. As we have seen in Sec.\,\ref{sec:symuniversality}, symmetric product orbifolds also satisfy this condition \cite{Pakman:2009zz}.

\paragraph{Growth of light correlators in medium states.} The third condition regards correlators of light operators computed in medium states. The condition is that such correlators grow at most polynomially in the central charge, that is, for an operator $\bfO_A\in M$, and light operators $\bfO_1,\dots,\bfO_n$, we have
\eq{
\bra{A}\bfO_1(x_1)\cdots\bfO_n(x_n)\ket{A}=\mathcal{O}\left(c^p\right)\,,
}
for some positive $p$. From an AdS$_3$ point of view, this condition states that there are some solutions in which light fields can take macroscopic values, but do not contain enough energy to create a black hole.

It is instructive to analyze this condition in detail for one-point functions of medium states in symmetric product orbifolds. To do so we have to understand what medium states look like in this setup. There are basically two ways to create states that are not light. First, one can pick a very heavy state in the seed and symmetrize it. Such a state can never give rise to an exponential growth in $c=Nc_\text{seed}$. In fact, in that case, the relevant ingredient is the seed OPE coefficient $\bra{H}\bfOs\ket{H}$, where $\Delta_{H}\gg c_{\text{seed}}$. This OPE coefficient does not scale exponentially with $N$. If we furthermore average over the heavy operators of the theory, it is exponentially suppressed \cite{Kraus:2016nwo,Collier:2019weq}. Hence, medium states created in this way do not lead to exponential growth with $N$. The second way one can create a medium state is by exciting many copies of the orbifold. In other words, by considering multi-trace states with many traces. The OPE coefficients of (untwisted versions of) such operators are derived in \cite{Belin:2017nze}
\eq{
C_{123}\coloneqq\left\langle \mathop{:}\!\bfO^{K_1}\!\mathop{:}\mathop{:}\!\bfO^{K_2}\!\mathop{:}\mathop{:}\!\bfO^{K_3}\!\mathop{:}\right\rangle=\frac{\sqrt{K_1!K_2!K_3!}}{\left(\frac{K_1+K_2-K_3}{2}\right)!\left(\frac{K_1-K_2+K_3}{2}\right)!\left(\frac{-K_1+K_2+K_3}{2}\right)!}+\cdots\,.
}
Here we only kept the leading order in $N$. When we take two of the operators to have trace $M$, and the other one of trace $L$, we find
\eq{
C_{123}\sim\frac{M!\sqrt{L!}}{\left(\left(\frac{L}{2}\right)!\right)^2\left(M-\frac{L}{2}\right)!}\,.
}
When we take $M$ to scale with $N$ this leads to a polynomial growth in $N$ with power $L/2$. Including (finitely many) more light insertions cannot increase the scaling to be exponential and thus the condition is satisfied for the operators considered in this simple example: a symmetrized version of the $M$-th power of a single-trace untwisted operator. From the more general expressions for the OPE coefficients that also include twisted sector operators of \cite{Belin:2017nze} a similar argument suffices to show that the condition is met in symmetric product orbifolds.

\paragraph{Large central charge expansion of light thermal correlators.
}

The final condition of \cite{Kraus:2017kyl} regards the large central charge expansion of the light contribution to thermal correlators. We denote the light contribution to the thermal correlator of light operators $\bfO_i$ as
\eq{
G_n^L(\beta)\coloneqq
\frac{e^{\beta c/12}}{Z(\beta)}\sum_Le^{-\beta\Delta_L}\bra{L}\bfO_1\cdots\bfO_n\ket{L}\,.
}
For low temperatures, $\beta>2\pi$, we perform the large-$c$ expansion using large-$N$ factorization, and then take $\Delta_c\rightarrow\infty$ to get
\eq{
G_n^{(L)}(\beta)=\sum_{k=0}^\infty\frac{1}{c^{k/2}}G_n^{(L,k)}(\beta)\,.
}
The condition then states that $G_n^{(L,k)}(\beta)$ is well defined in the limit $\Delta_c\rightarrow\infty$.

Symmetric product orbifolds also satisfy this last condition. This can be seen from the fact that in the large-$N$ limit, any correlator can be systematically expanded in $1/N$, and at every order the contribution is known in terms of seed quantities \cite{Belin:2015hwa}.
\\

We can thus conclude that symmetric product orbifolds satisfy all conditions of \cite{Kraus:2017kyl}. Hence, the thermal one- and two-point functions can be written as \eqref{eq:kss1pt} and \eqref{eq:kss2pt} respectively. The results of Sec.\,\ref{sec:thuni} complement the results of this appendix nicely. In the main text, we are able to compute the explicit leading order behavior of the thermal one- and two-point functions of untwisted operators in symmetric product orbifolds. The expressions in this appendix are more general. They are true for all light operators and are valid up to exponentially suppressed corrections in $N$ and the cutoff. Part of the price for this generality is the less explicit form for the thermal correlation functions in terms of a sum over light contributions. 

Although the terms in the sum bear resemblance to bulk Witten diagrams in an AdS$_3$ geometry (thermal or BTZ depending on the temperature), it is not clear that the leading order answer for, say the thermal two-point function of untwisted operators, is given by the two-point function of light fields in the BTZ geometry for $\beta<2\pi$. Furthermore, it is a priori not obvious how the sum over images is encoded in the sum over light operators. For the first two non-trivial images, this was worked out by \cite{Alexunpublished}. The trivial image, \ie, the term in the sum in \eqref{eq:finalth} with $j=0$ can be extracted from \eqref{eq:kss2pt} by considering $A$ to be the identity operator. In that case, the OPE coefficient is only nonzero if $B=\partial^k\bar{\partial}^{\bar{k}}\bfO$. The expression for the OPE coefficients $C_{I\bfO\partial^k\bar{\partial}^{\bar{k}}\bfO}$ can be summed over $k$ and $\bar{k}$ and gives rise to the $j=0$ term of \eqref{eq:finalth}. To also obtain the first non-trivial image, that is, the $j=1$ term of \eqref{eq:finalth}, one has to additionally take into account terms with $A=\partial^{k_1}\bar{\partial}^{\bar{k}_1}\bfO$ and $B=\partial^{k_2}\bar{\partial}^{\bar{k}_2}\mathop{:}\!\bfO\partial^{k_3}\bar{\partial}^{\bar{k}_3}\bfO\!\mathop{:}$. Presumably, in this way one can systematically disentangle the entire sum over images from \eqref{eq:kss2pt}. It would be very interesting to show how this works explicitly. One hurdle that one has to overcome is that for the higher and higher images, one needs to understand the OPE coefficients of higher and higher trace operators.

\bibliographystyle{ytphys}
\bibliography{ref}
\end{document}